\documentclass[review,number,sort&compress]{elsarticle}

\usepackage{multirow} 
\usepackage{graphicx}
\usepackage{subfigure}
\usepackage{amsmath}
\usepackage{amssymb}
\usepackage{textcomp }
\usepackage{ulem}

\usepackage{lineno}

\usepackage{geometry}
\newcommand{\nubb}{$\nu\beta\beta$}
\newcommand{\up}[1]{$^{#1}$}
\newcommand{\power}[1]{$^{#1}$}

\newcommand{\gerda}{{\sc Gerda}}
\newcommand{\mage}{{\sc MaGe}}
\newcommand{\ger}{\up{76}Ge}
\newcommand{\tho}{\up{228}Th}
\newcommand{\cd}{$\cdot$}
\newcommand{\cts}{cts/(keV\cd kg\cd yr)}

\newcommand{\dir}{./}

\journal{NIMA}

\begin{document}

\begin{frontmatter}



\title{Monte Carlo Studies and Optimization for the Calibration System of the \gerda\ Experiment}


\author[zh]{L. Baudis} 
\author[zh,lngs]{A.D. Ferella}
\author[zh]{F. Froborg\corref{cor1}}
\ead{francis@froborg.de}
\author[zh,urbana]{M. Tarka}

\cortext[cor1]{Corresponding author}
\address[zh]{Physics Institute, University of Zurich, Winterthurerstrasse 190, 8057 Z\"urich, Switzerland}
\address[urbana]{Physics Department, University of Illinois, 1110 West Green Street, Urbana, IL 61801, US}
\address[lngs]{INFN – Laboratori Nazionali del Gran Sasso, 67010 Assergi, Italy}

\begin{abstract}
The GERmanium Detector Array, \gerda, searches for neutrinoless double beta decay in \ger\ using bare high-purity germanium detectors submerged in liquid argon. For the calibration of these detectors $\gamma$ emitting sources have to be lowered from their parking position on top of the cryostat over more than five meters down to the germanium crystals. With the help of Monte Carlo simulations, the relevant parameters of the calibration system were determined. It was found that three \tho\ sources with an activity of 20\,kBq each at two different vertical positions will be necessary to reach sufficient statistics in all detectors in less than four hours of calibration time. These sources will contribute to the background of the experiment with a total of  \linebreak $(1.07\pm0.04(\text{stat})^{+0.13}_{-0.19}(\text{sys}))\times 10^{-4}$\,\cts) when shielded from below with 6\,cm of tantalum in the parking position.

\end{abstract}


\end{frontmatter}



\section{Introduction}

Experiments with solar, atmospheric and accelerator neutrinos as well as with reactor and accelerator antineutrinos have provided compelling evidence for neutrino mixing and for non-zero neutrino masses [see e.g.~\cite{nakamura12} and references therein]. The observed phenomena are neutrino oscillations in vacuum and flavour transformation in matter. Both phenomena depend on the three neutrino mixing angles, which have been experimentally determined\cite{kamland08,minos08,theta13comb11}. Moreover, the former depends, at the first order, on the absolute value of the larger squared neutrino mass difference, the latter on the smaller one in absolute value and sign~\cite{fogli12}. While these quantities are known, the sign of the larger mass difference, the absolute value of the masses and the CP violating phase (or phases) remain to be measured. 
\\
These experiments are not sensitive to the nature of neutrinos, namely whether neutrino and antineutrino are the same (Majorana) or different (Dirac) particles. The observation of the neutrinoless double beta decay (0\nubb) would prove that neutrinos have a Majorana component and that consequently lepton number is not conserved. The measured half-life will provide information on the Majorana effective electron neutrino mass $m_\text{ee}$, using the calculated nuclear matrix elements. Several experiments have searched for 0\nubb, but none provided positive uncontroversial evidence~\cite{avignone07, bilenky10, rodejohann11, gomez11, schwingenheuer12,bilenky12}. Current experimental upper limits on $m_\text{ee}$ are in the range of $0.3-1.0$~eV. The most stringent upper limits on the half-lives have been established by experiments on \ger\ (with enriched HPGe diodes) (Heidelberg-Moscow~\cite{klapdor01HdM} and IGEX~\cite{igex02}), on 130Te, with natural tellurium oxide bolometers (CUORICINO~\cite{cuoricino2012}), on 100Mo with enriched metal foils (NEMO3~\cite{nemo11,nemo10}) and on 136Xe with a liquid TPC (EXO-200~\cite{exo12}) and dissolved in liquid scintillator (KamLAND-ZEN~\cite{kamland-zen12}).

The \textsc{Ger}manium \textsc{D}etector \textsc{A}rray, \gerda, aims to significantly improve these results using bare HPGe detectors submerged in liquid argon~\cite{gerdaOverview}. To calibrate these detectors, radioactive sources are used. This paper presents the Monte Carlo studies necessary to determine the optimum type of source, its activity, calibration position and background contribution. It is structured as follows: Section~\ref{sec:gerda} gives a brief overview of the \gerda\ experiment, while section~\ref{sec:calsys} specifies the requirements on the calibration system. The general setup of the Monte Carlo simulations is described in section~\ref{sec:mcs}; the results are shown in section~\ref{sec:results}. In section~\ref{sec:background}, the background contribution from the calibration sources is determined. Section~\ref{sec:conclusion} presents the conclusions.


\section{The \gerda\ Experiment}\label{sec:gerda}
\begin{figure}[t]
	\centering
	\includegraphics[width=.35\textwidth]{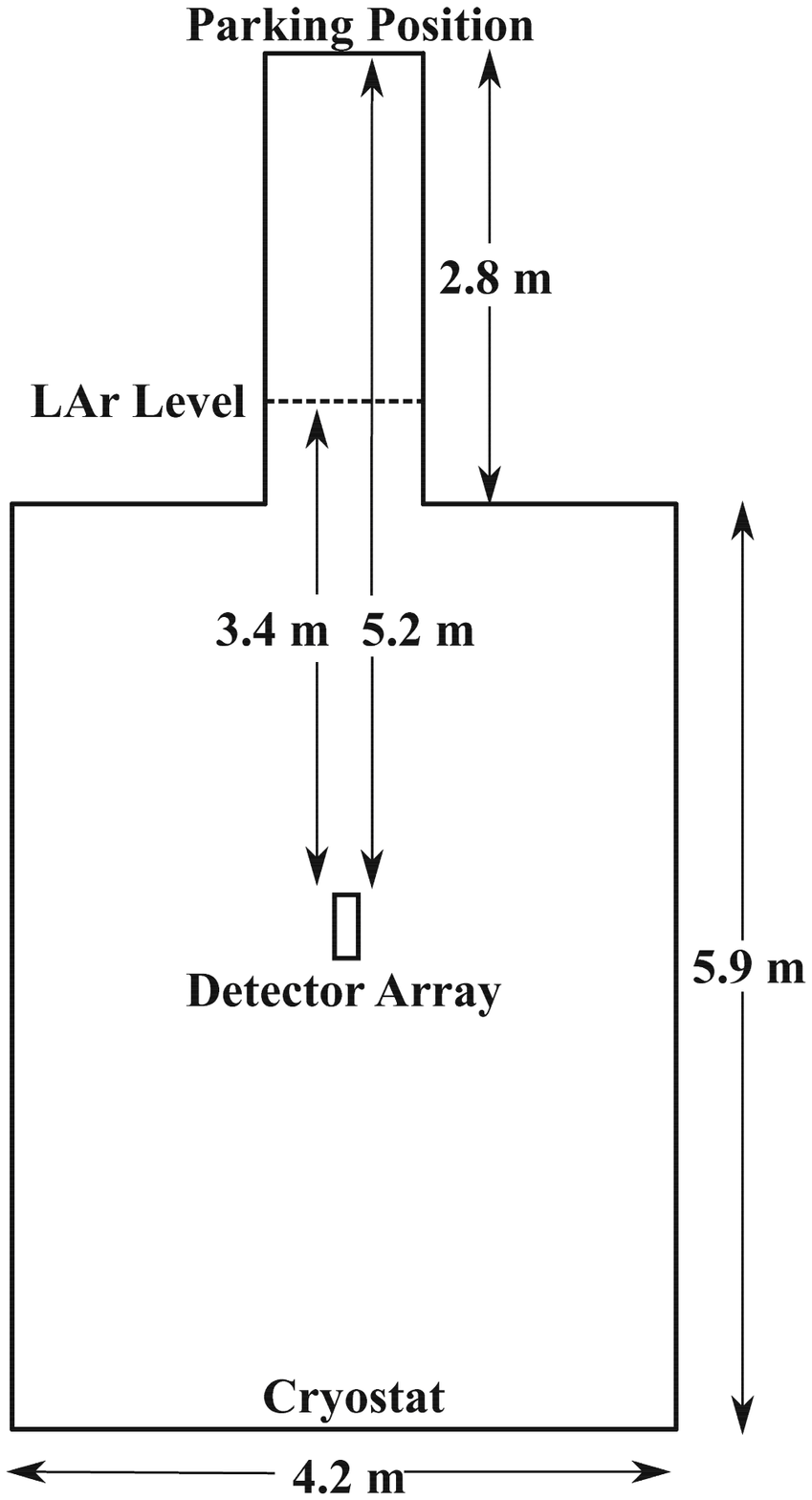}
	\caption{Sketch of the \gerda\ cryostat. Indicated are the positions of the detector array as well as the parking position for the calibration sources.}
	\label{fig:gerda}
\end{figure}

The \gerda\ experiment is searching for the 0\nubb\ decay in \ger\ which has a half life above $10^{25}$\,yr. To detect this rare process it is necessary to shield the detectors from the radioactivity of the surrounding materials as well as cosmic radiation. For the latter the underground location of the Laboratori Nazionali del Gran Sasso (LNGS) of INFN was chosen, where 3800\,mwe (average for the muon angular distribution and the mountain profile) of rock shields the experiment~\cite{lngsShielding11}. Furthermore, only materials with a very low intrinsic radioactivity were used to build the experiment, and the detectors are submerged nakedly into liquid argon (LAr). The argon acts simultaneously as cooling liquid and shielding material. If equipped with photomultipliers, it can be used as an active veto~\cite{larveto07}. A sketch of the cryostat with all relevant dimensions is shown in figure~\ref{fig:gerda}. The argon cryostat is surrounded by a 10\,m diameter water Cerenkov veto, which further shields the detectors.

The experiment is foreseen to proceed in two phases. Phase I is using reprocessed HPGe detectors from the Heidelberg-Moscow and IGEX experiments enriched to 86\% in \ger\ with a total mass of 17.7\,kg together with natural HPGe detectors from the Genius Test Facility~\cite{gtf}. The goal is to improve upon current limits on the 0\nubb\ sensitivity with an exposure of 15\,kg\cd~yr and a background of $<10^{-2}$\,\cts). In case no events will be observed above background, a half life limit of $T_{1/2} > 2.2\times 10^{25}$\,yr can be established, resulting in an upper limit for the effective neutrino mass of $m_\text{ee}<0.23-0.39$\,eV~\cite{smolnikov10}. Phase II will use in addition newly developed broad-energy germanium detectors~\cite{bege} enriched in \ger, and is foreseen to run for a total exposure of 100\,kg\cd yr. With an aimed background of $<10^{-3}$\,\cts) a half life sensitivity of $T_{1/2} >15\times 10^{25}$\,yr and a corresponding effective neutrino mass of $m_\text{ee}<0.09-0.15$\,eV~\cite{smolnikov10} can be reached.

\section{The Calibration System}\label{sec:calsys}
For energy as well as pulse shape calibrations of the HPGe detectors radioactive sources are used. These sources are staying inside the cryostat to prevent radioactive contamination of the liquid argon due to reentry. The parking position of the sources during physics runs is on top of the cryostat which is indicated in figure~\ref{fig:gerda}. During a calibration run the sources are lowered from the top of the cryostat down to the detectors. The entire calibration including the movement of the sources should not take more than 4 hours to reduce dead time for physical measurements. On average, a calibration run is performed once a week.

This requires that the activity of the calibration sources should not be too high, such as to enable sufficient shielding while in parking position, to ensure a low background contribution. On the other hand, the activity has to be high enough to guarantee that the total calibration will be performed in a reasonable amount of time. To reach this goal Monte Carlo simulations were used to optimize parameters such as activity as well as the number of sources and their exact positions during calibration runs. Furthermore, the resulting background contribution was determined together with the necessary shielding of the source(s). This paper focuses on Phase~I of the experiment although most of the results can be transferred to Phase II as well. Further details can be found in \cite{myThesis}.


\section{Monte Carlo Simulations}
\label{sec:mcs}
\begin{figure}[t]
	\center
	\subfigure[Technical drawing of the lit of the cryostat showing the position of the four detector strings D1-4 and the three calibration sources S1-3.]{
		\includegraphics[width=.4\textwidth]{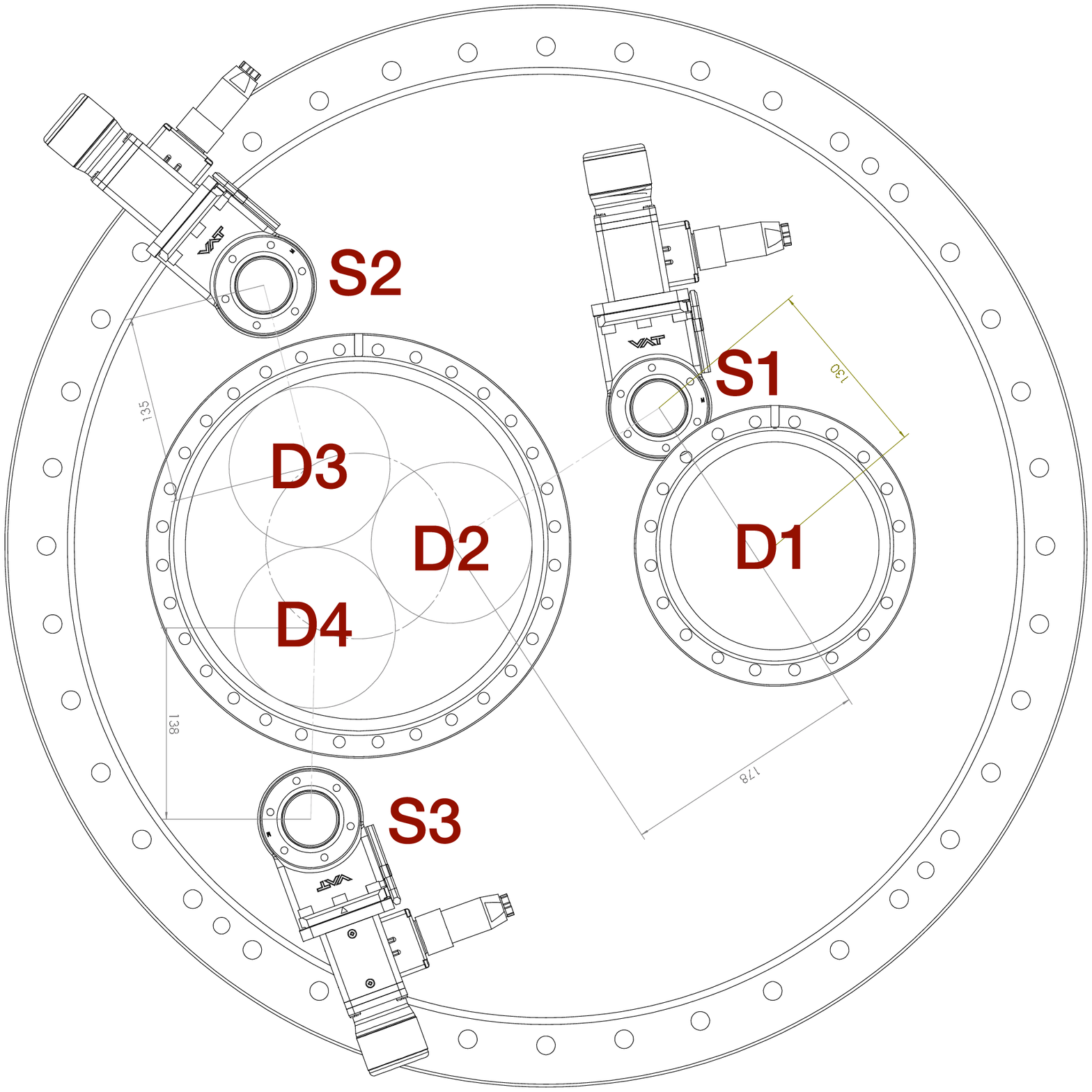}\label{fig:p1top}
	}
	\qquad
	\subfigure[Side view of the detector array. Shown in red are the detectors with their different heights, shown in grey are the three calibration sources with their absorbers.]{
		\includegraphics[width=.36\textwidth]{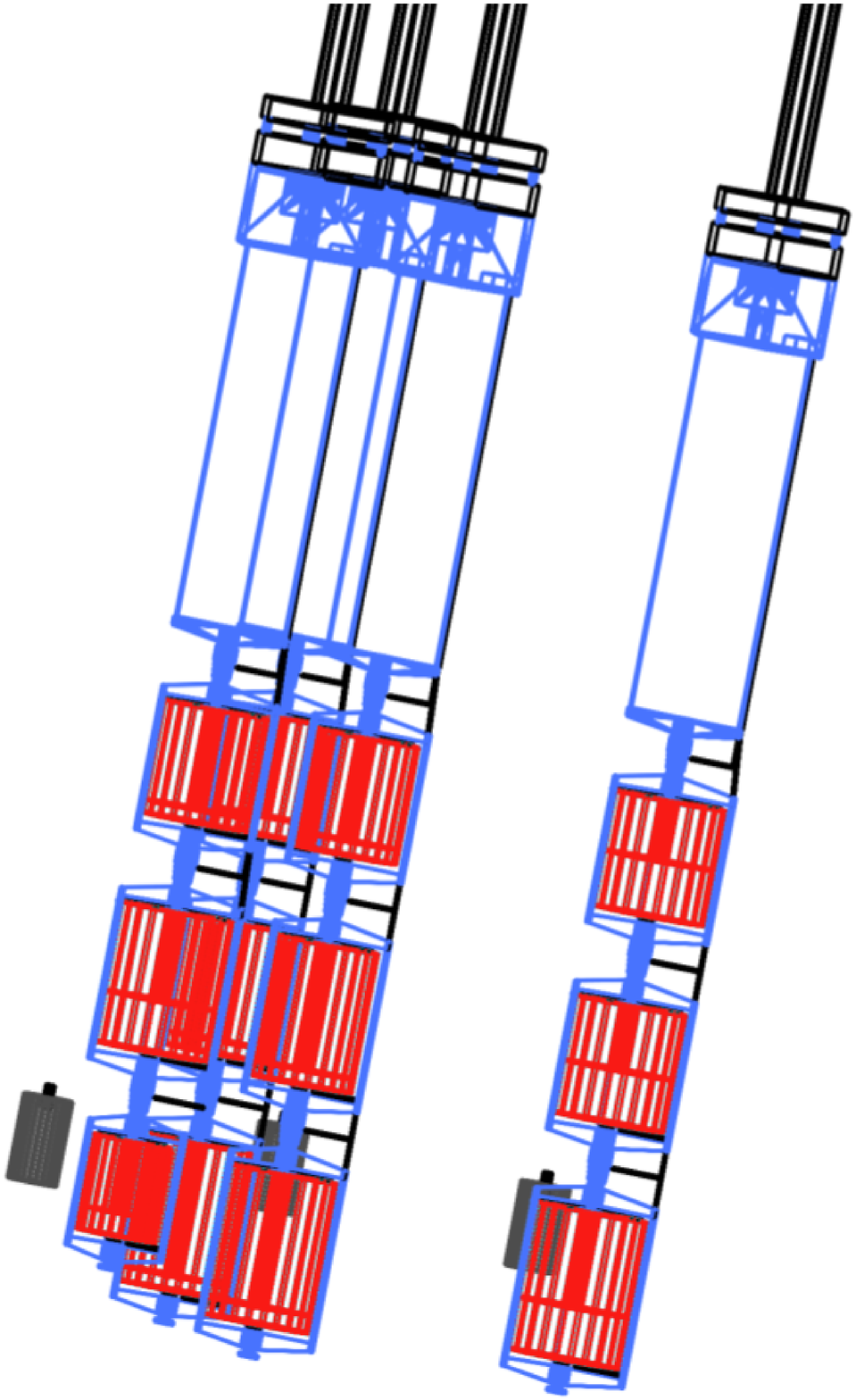}\label{fig:p1side}}
	\caption{Configuration of detectors and calibration sources during \gerda\ Phase I.}\label{fig:p1config}
\end{figure}

The Monte Carlo simulations were performed in the \mage~\cite{mage12} framework, a package based on {\sc Geant4}~\cite{geant403, geant406}, which simulates the geometry of the entire \gerda\ experiment as well as all relevant physics processes. If not indicated differently, the following setup was used: The 8 enriched and 4 non-enriched detectors were placed according to the original Phase~I configuration shown in figure~\ref{fig:p1config}. They are arranged in four strings D1-4 with three detectors per string (see figure~\ref{fig:p1top}). The detectors are different in size (see figure~\ref{fig:p1side}) as shown in table~\ref{tab:detarray}; they are aligned to the top edge of the top detectors.

Calibration sources at three different positions S1-3 were simulated with a total of $10^8$ decays; their position relative to the detector array can be seen in figure~\ref{fig:p1top}. The source encapsulation and intended composition were included. If the daughter is not stable, the full decay chain was simulated with a pause of 100 microseconds between the decay of the different isotopes of the chain to prevent unrealistic pile-up events.

The resulting energy spectra were folded with the expected energy resolution of the detectors which was determined from first test data taken in May 2009 with a \power{60}Co and a \power{232}Th source. Table~\ref{tab:eres} shows the energy resolution obtained at different energies for one detector.

\begin{table}[t]
	\centering
	\caption{Configuration of the detector array including the height $h$ of the different detectors. D1-4 refer to four strings with three detectors each. D4.3 is by far the smallest detector and thus is special.}
	\begin{tabular}{c|cccc}
	$h$ [mm] & D1 & D2 & D3 & D4\\\hline\hline
	1 & 84 & 84 & 85 & 85\\
	2 & 86 & 101 & 94 & 105\\
	3 & 107 & 108 & 106 & 68\\\hline
	\end{tabular}
	\label{tab:detarray}
\end{table}

\begin{table}[t]
	\centering
	\caption{Energy resolution for different energies taken with a \power{60}Co and a \power{228}Th source and one of the \gerda\ Phase I detectors during testing in May 2009.}
	\begin{tabular}{l|cccccccc}
	$E$ [keV] & 511 & 586 & 911 & 969 & 1173 & 1333 & 2104 & 2615\\\hline
	$\sigma$ [keV] & 1.73 & 1.35 & 1.42 & 1.50 & 1.47 & 1.51 & 2.12 & 1.78
	\end{tabular}
	\label{tab:eres}
\end{table}


\section{Results}
\label{sec:results}

\subsection{Type of Source}
\label{sec:source}
Since the $Q$-value of \ger\ is at $2039$\,keV~\cite{douysset01}, the 5\,keV region around this value was chosen as the region of interest (ROI). Therefore, several lines in the energy range up to 2.5\,MeV together with at least one line close to the $Q$-value are necessary for the energy calibration. Since the signal appears as a single-site event (SSE) in the detector, all multi-site events (MSE) can be rejected as background which is typically done in an offline pulse shape analysis. To calibrate the corresponding parameters, a clear double escape peak (DEP) is needed as a sample of SSE. This requires a strong full energy peak (FEP) well above 2\,MeV, such that the probability of pair production is sufficiently high, with both annihilation photons escaping. A peak to background ratio of at least 2:1 was considered sufficient in this case. A full energy peak close to the DEP would be an asset because it would allow the comparison of a sample of SSE and a sample of MSE without the influence of the energy dependent energy resolution of the detectors. Since the source will stay in the cryostat, a half life of at least several months is required.

Possible emission of $\alpha$ particles by the source is an important issue because they might produce neutrons in ($\alpha$,n) reactions. The probability of ($\alpha$,n) reactions depends on the energy of the $\alpha$ particles and the threshold energy of the material close to the source. Neutrons can contribute to the background in the ROI due to scattering or neutron capture, the latter resulting in radioactive isotopes which might emit photons or betas with an energy close to the ROI. 

\begin{figure}[t]
	\centering
	\subfigure[\power{56}Co.]{
		\includegraphics[width=.315\textwidth]{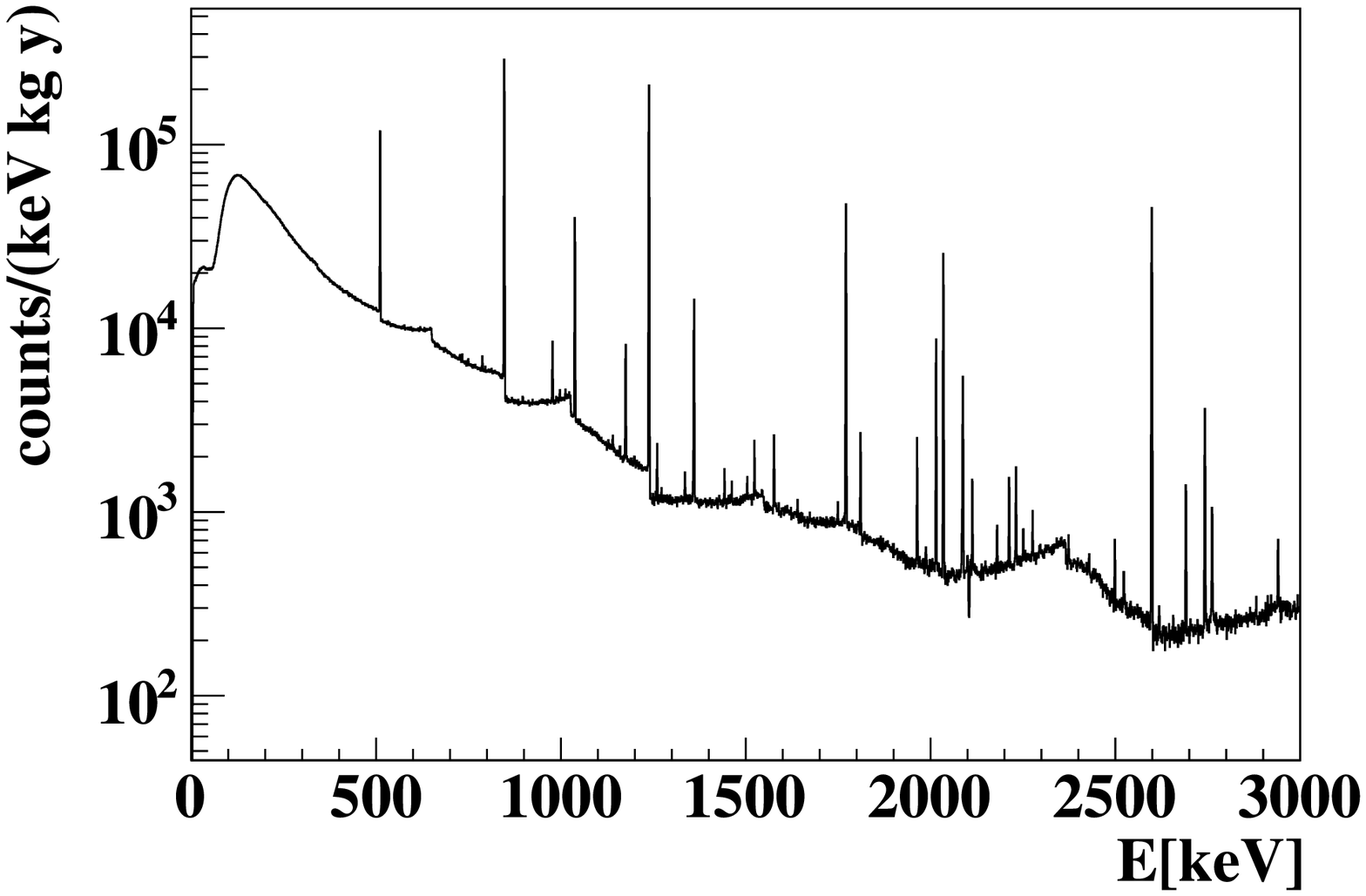}
		\label{fig:co56}
	}\hfill
	\subfigure[\power{228}Th.]{
		\includegraphics[width=.315\textwidth]{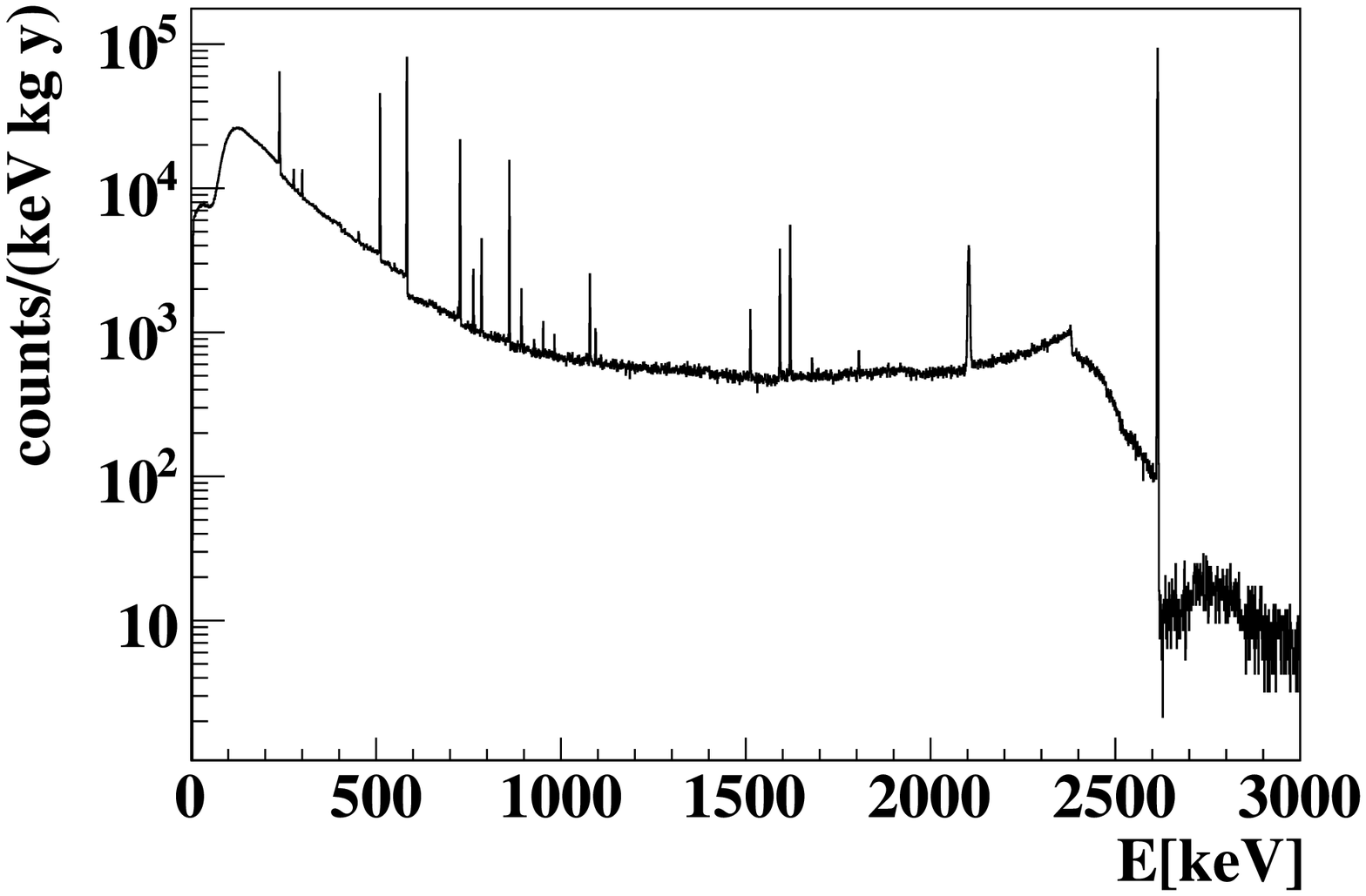}
		\label{fig:th228}
	}\hfill
	\subfigure[\power{238}U.]{
		\includegraphics[width=.315\textwidth]{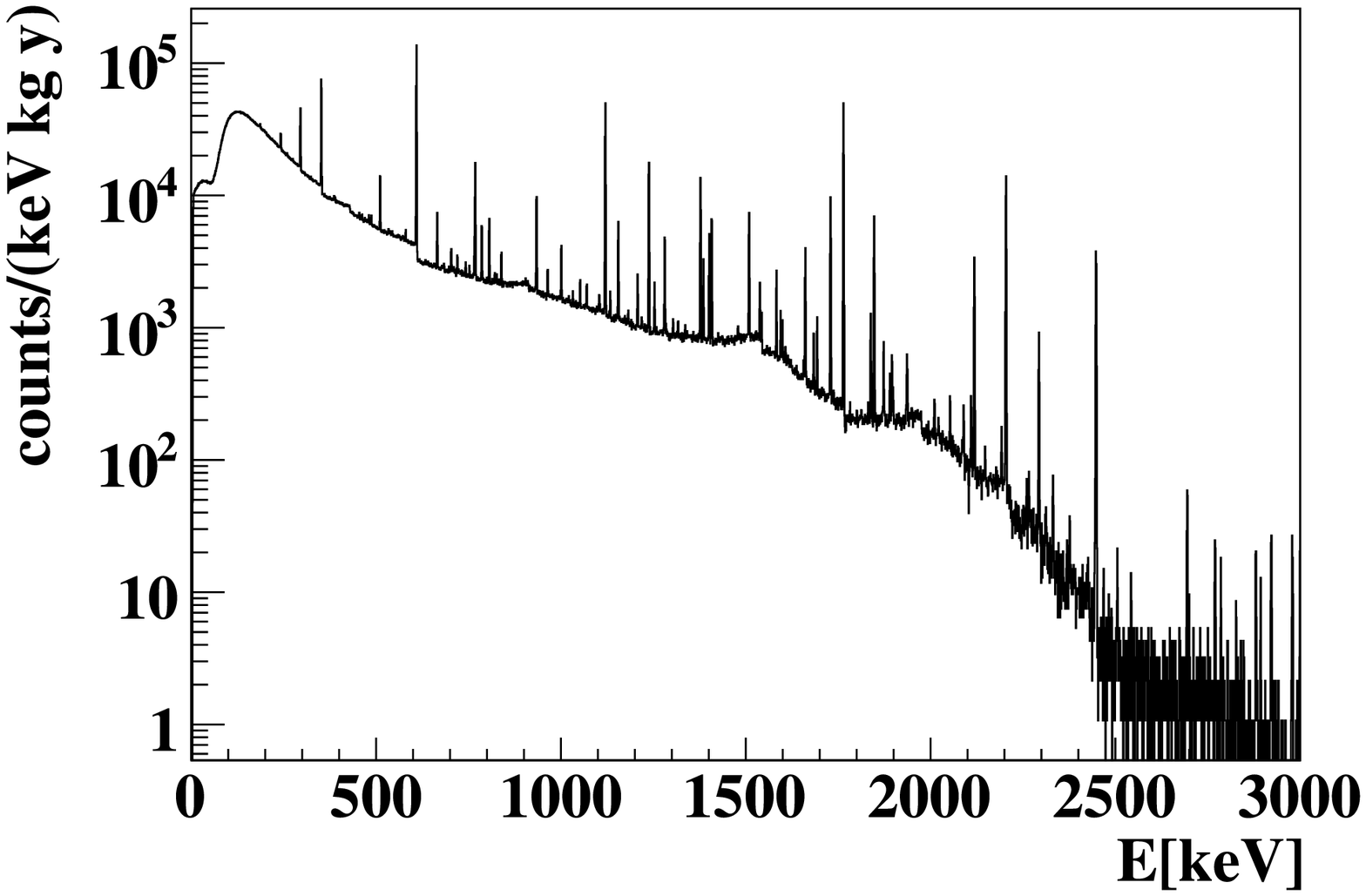}
		\label{fig:u238}
	}
	\caption{Comparison of different possible calibration sources. Shown is the sum of all entries in all detectors normalized to a source activity of $3\times20$\,kBq (the motivation for this value is given in sections \ref{sec:position} and \ref{sec:activity}) and the total detector mass.}
	\label{fig:type}
\end{figure}

Taking these requirements into account, three possible calibration sources are considered: \power{56}Co, \tho\ and \power{238}U; table \ref{tab:type} summarizes their relevant characteristics. These sources were studied in Monte Carlo simulations, positioned as explained in section~\ref{sec:position}. Figure~\ref{fig:type} shows the sum of the energy spectra of all detectors for the different sources normalized to the same activity, time and detector mass.

\begin{table}[t]
	\centering
	\caption{Relevant characteristics for the three possible calibration sources. $E_\text{max}$ gives the maximum energy photons are emitted with. The energy of the line closest to $Q=2.039$~MeV with an intensity above 1~\% is given. It is also indicated if the sources emits any $\alpha$ particles.}
	\begin{tabular}{l|ccc}
	 & \power{56}Co & \tho\ & \power{238}U\\\hline\hline
	$T_{1/2}$ & 77\,d & 1.9\,yr & $4\times 10^9$\,yr\\
	$E_{\gamma,\text{max}}$ [MeV] & 3.5 & 2.6 & 2.4\\
	$E_\gamma$ closest to $Q$-value [MeV] & 2.035 & 2.104 & 2.204\\
	$E_{\alpha,\text{max}}$ [MeV] & - & 7.7 & 8.8\\\hline
	\end{tabular}
	\label{tab:type}
\end{table}

As expected, all three sources show several well pronounced lines in the relevant energy region. The advantage of \power{56}Co is that it is not emitting any $\alpha$ particles. Furthermore, \power{56}Co has with $E_\gamma = 2.035$\,MeV a FEP very close but still distinguishable from the $Q$-value. The disadvantages of \power{56}Co are its short half life of $T_{1/2} = 77$\,d and no DEP with sufficient statistics for a pulse shape calibration. The next source considered is \power{238}U with a half life of $T_{1/2} = 4\times 10^9$\,yr and a FEP close to the $Q$-value at $E_\gamma=2.204$\,MeV. Unfortunately is does no show a DEP with sufficient statistics. Moreover it is a possible neutron producer due to spontaneous fission or emitted $\alpha$ particles with energies up to $E_\alpha = 8.8$\,MeV. The last source considered is \tho\ with a half life of $T_{1/2} = 1.9$\,yr and a peak close to the $Q$-value at $E_\gamma = 2.104$\,MeV. This is the single escape peak coming from the \power{208}Tl line at $E_{\text{Tl}}=2.615$\,MeV. It also results in a reasonable DEP at $E_{\text{DEP}}  = 1.593$\,MeV with a peak to background ratio of 2:1. With the \power{212}Bi line at $E_{\text{Bi}}=1.621$\,MeV there is a FEP very close to the DEP which makes it ideal for pulse shape calibration. The disadvantage of the source is that it emits $\alpha$ particles which can produce neutrons in the surrounding material. The implications will be discussed in section \ref{sec:nback}. Nonetheless, it is the best option for the experiment and is therefore used for the actual calibration measurements\cite{gerdaOverview, gerda2nu2013}.

\subsection{Positioning}
\label{sec:position}

\begin{figure}[t]
	\centering
	\includegraphics[width=.4\textwidth]{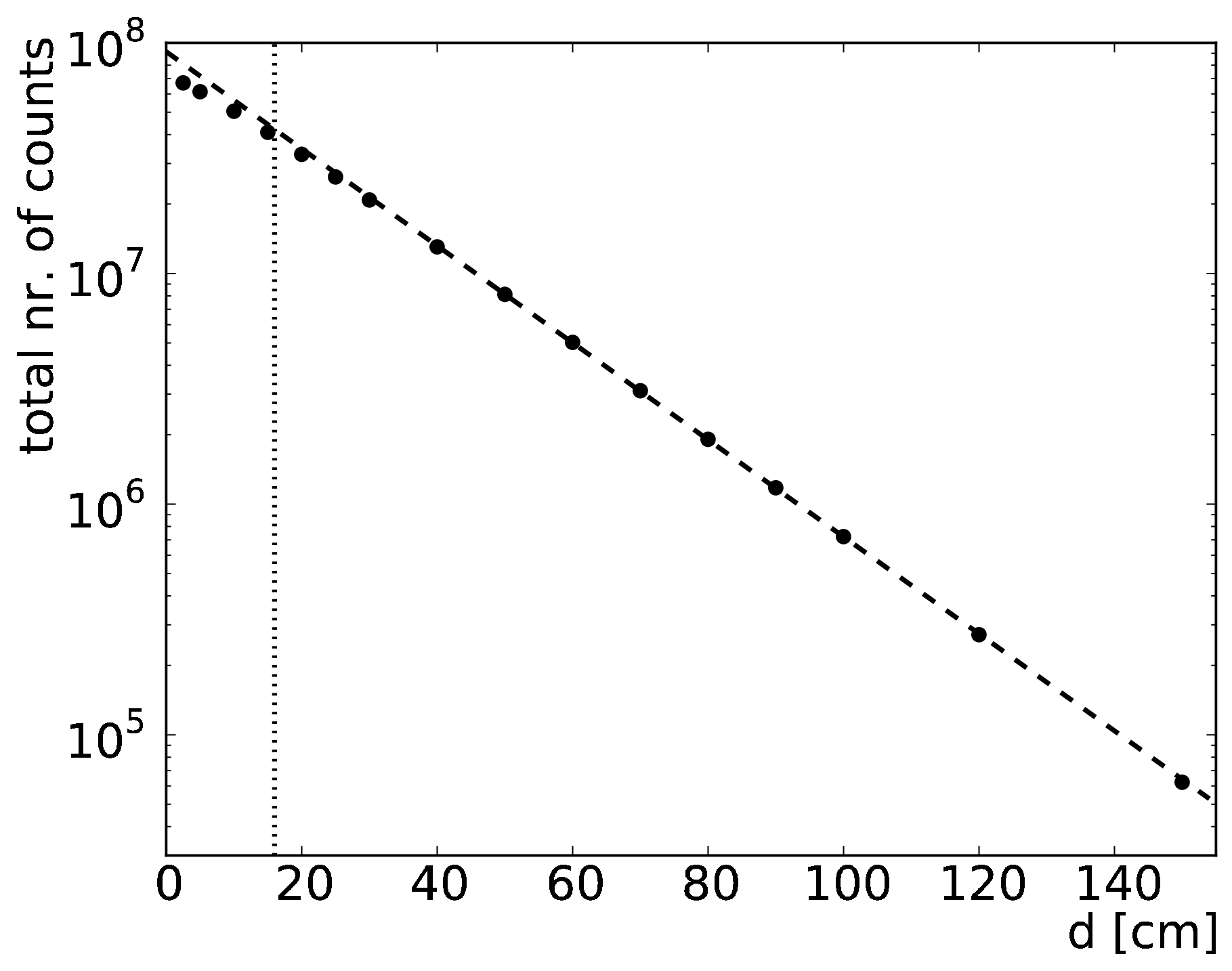}
	\caption{Total  number of counts in one detector for different distances to the source. An exponential decay was fitted to the data points to determine the mean free path of 2.6~MeV $\gamma$'s in LAr. The dotted line marks the average distance between source and detector as it is realized in \gerda\ Phase I.}
	\label{fig:disttest}
\end{figure}

\begin{figure}[p]
	\centering
	\subfigure[String D1]{
	\includegraphics[width=.45\textwidth]{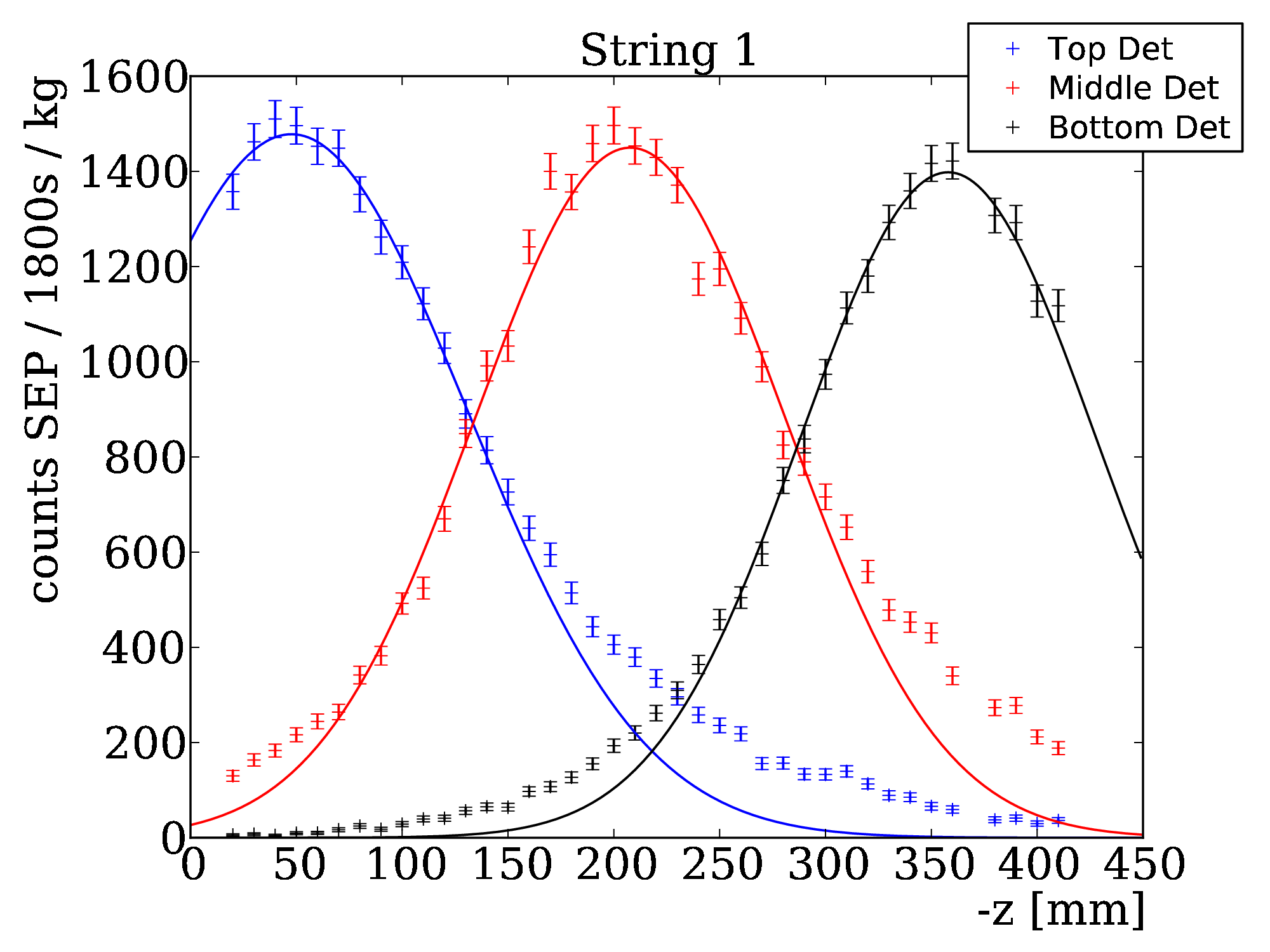}}
	\hfill
	\subfigure[String D2]{
	\includegraphics[width=.45\textwidth]{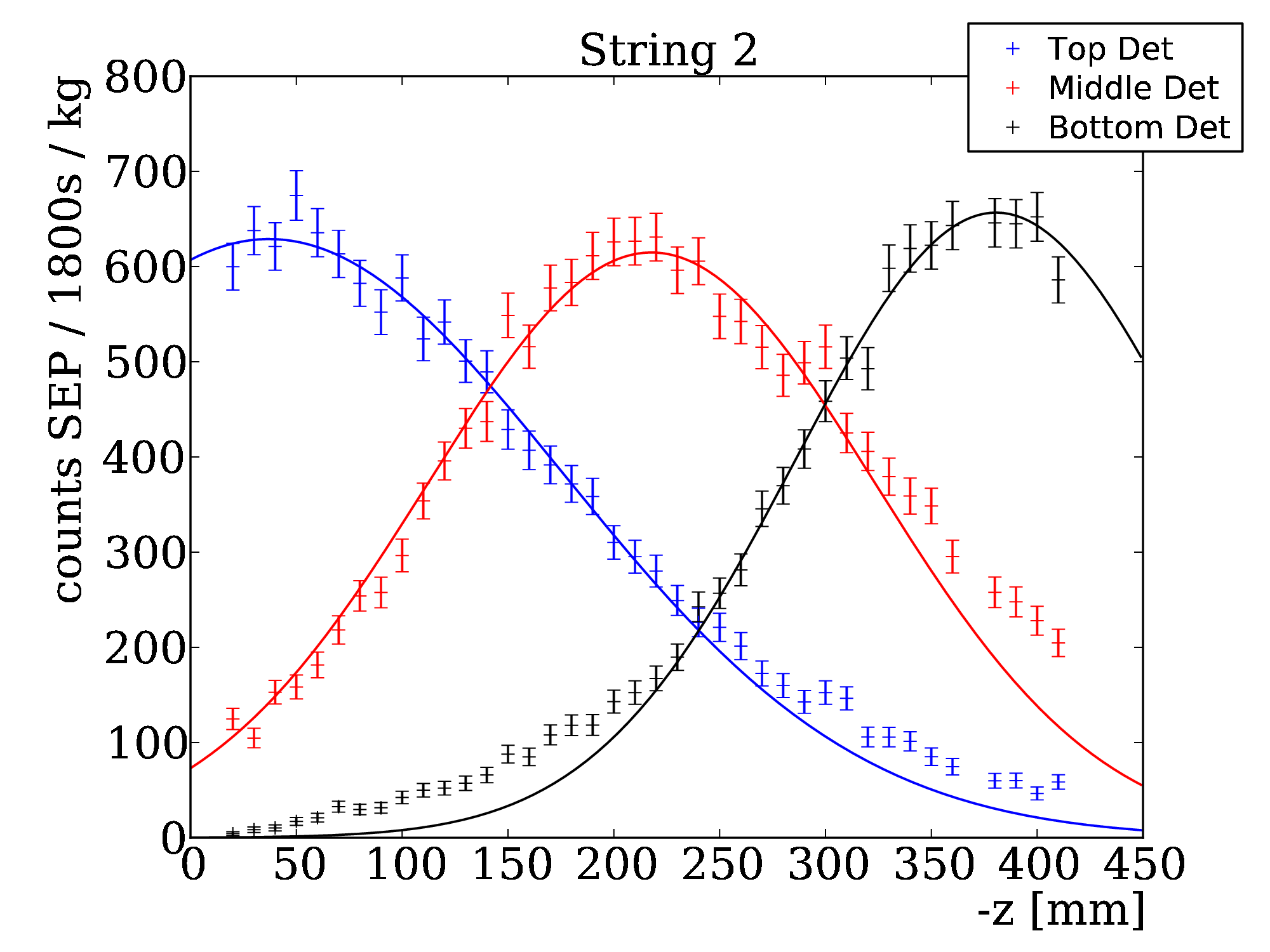}}\\
	\subfigure[String D3]{
	\includegraphics[width=.45\textwidth]{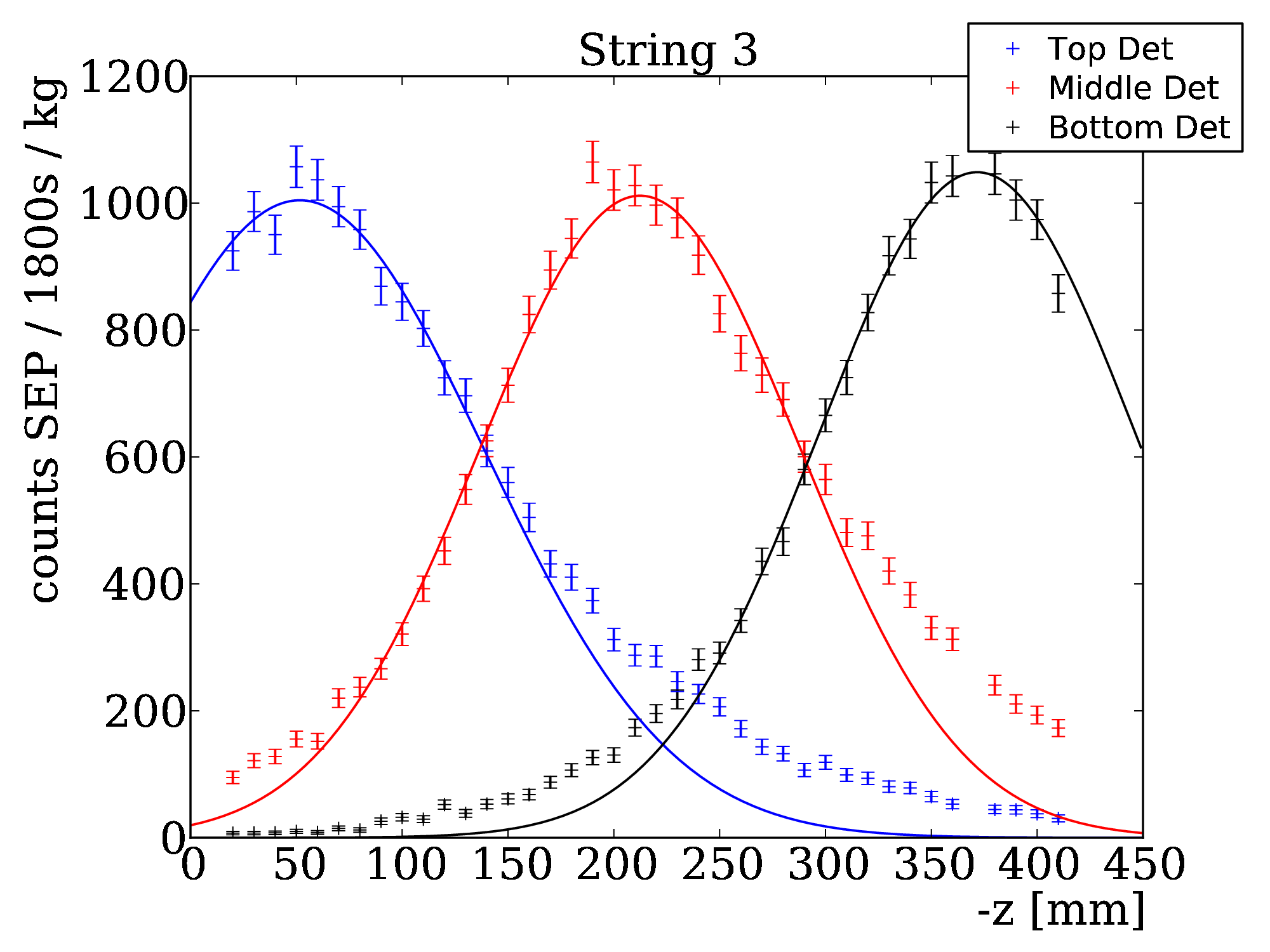}}
	\hfill
	\subfigure[String D4]{
	\includegraphics[width=.45\textwidth]{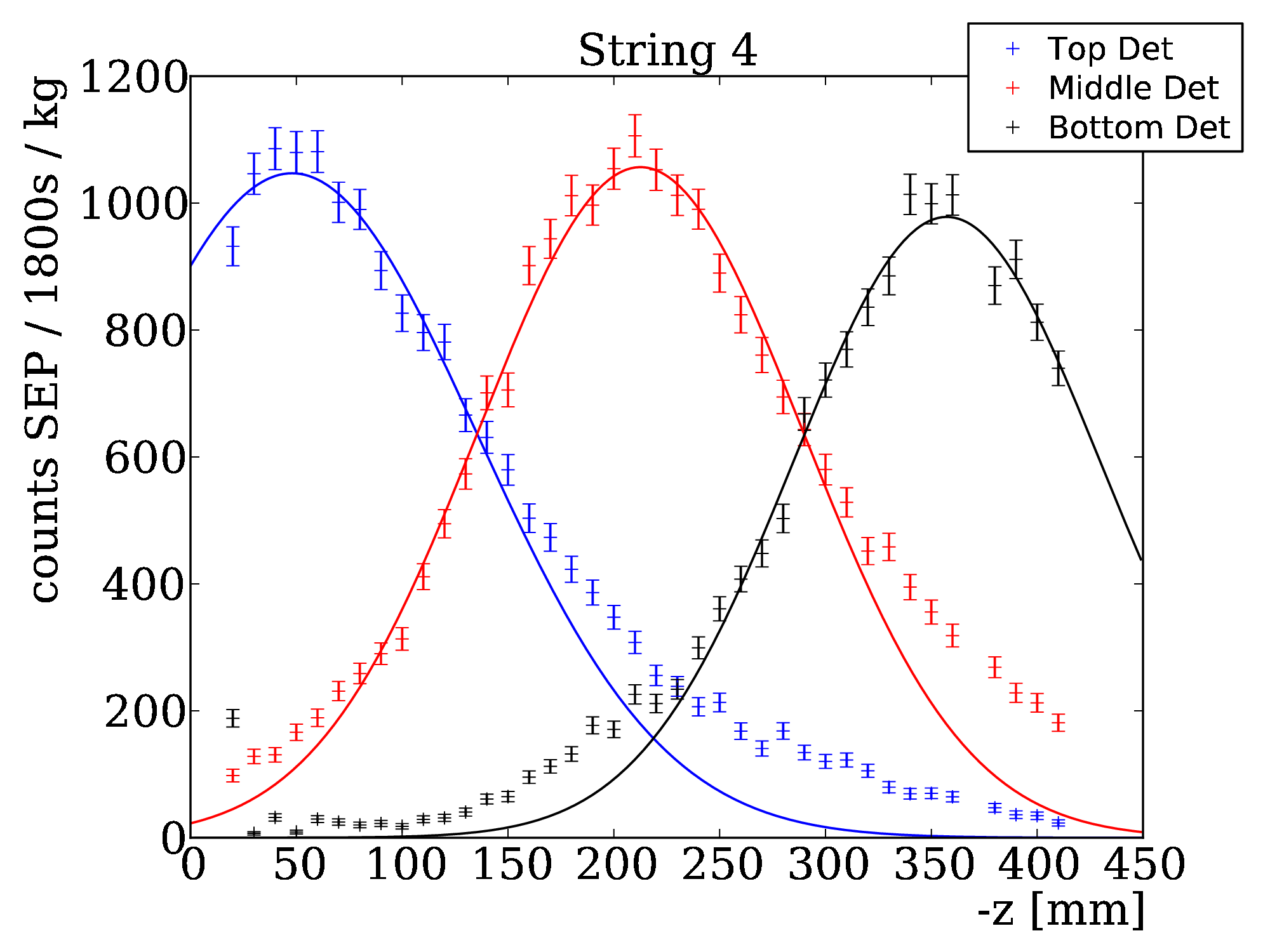}}\\
	\caption{Number of counts in the SEP in each detector normalized to a source strength of 20\,kBq, 1800\,s calibration time and the detector mass. Gaussian functions wer fitted to determine their peaks as positions with maximum count rate (3 calibration positions per source) as well as their intersections as optimum positions between two detectors (2 calibration positions).}
	\label{fig:optpos}
\end{figure}

\begin{table}[p]
	\centering
	\caption{Optimum vertical position $-z$ in [mm] relative to the top edge of the top detector of the calibration sources S1, S2 and S3 (see figure~\ref{fig:p1top}) for 2 and 3 vertical calibration positions per source.}
	\begin{tabular}{ccc|ccc}
	S1 & S2 & S2 & S1 & S2 & S3\\\hline
	50 & 50 & 50\\
	& & & 135 & 130 & 130\\
	210 & 215 & 215\\
	& & & 280 & 290 & 320\\
	360 & 380 & 365\\\hline
	\multicolumn{3}{c|}{3 Positions} & \multicolumn{3}{c}{2 Positions}
	\end{tabular}
	\label{tab:optpos}
\end{table}
\begin{table}[p]
	\centering
	\caption{Number of counts in SEP with $A=20$\,kBq per source in [counts/1800s] for 3 or 2 calibration positions as given in table \ref{tab:optpos}.}
	\begin{tabular}{r|cccc|cccc}
	 & D1 & D2 & D3 & D4 & D1 & D2 & D3 & D4\\\hline
	top detector & 5100 & 4000 & 4200 & 3800 & 2800 & 2600 & 2600 & 2200\\
	middle detector & 5300 & 4700 & 5100 & 5900 & 4400 & 4000 & 4100 & 4300\\
	bottom detector &  5500 & 4300 & 4800 & 1600 & 2800 & 2300 & 2500 & 1100\\\hline
	& \multicolumn{4}{c|}{3 Positions} & \multicolumn{4}{c}{2 Positions}
	\end{tabular}
	\label{tab:countssep}
\end{table}

Scattering processes in the liquid argon lead to a decrease of events in the detectors as well as a reduction of the peak to background ratio of the DEP with increasing distance between detector and calibration source. Figure~\ref{fig:disttest} shows the total number of counts in one detector for different distances to the source. It is evident that the source should be as close as possible to the detector. Additionally, the detector strings shield each other. Given the string configuration shown in figure~\ref{fig:p1config} it is obvious that more than one calibration source is necessary.

The limiting factor for the positioning of the sources is the lid of the cryostat which defines the (horizontal) $xy$ positions of the detector strings and the calibration sources. Predefined are the two flanges for the detector strings: One DN250 CF housing three detector strings and one DN160 CF housing one detector string (see figure~\ref{fig:p1top}). The smallest usable flange for the calibration sources are DN40 CF giving first limits for the smallest possible distance between source and detector. Space is also limited due to the fact that the pipes above the detector flanges have to be accessible to insert the detector strings as well as the calibration sources. Furthermore, the lowering system for the calibration sources, although reduced in size to an absolute minimum, needs some space. Taking all these factors into account, an optimum of three sources positioned horizontally as shown in figure~\ref{fig:p1config} is found. The average distance between source and detector of 16\,cm is marked in figure~\ref{fig:disttest}.

The determination of the best (vertical) $z$ positions for the sources is non-trivial due to the different heights of the detectors (see table~\ref{tab:detarray}), the self-shielding of the detectors and an absorber used to shield the sources in their parking position (see section \ref{sec:background}). Thus, Monte Carlo simulations were used with the same number of decays for each simulation and scanning through the $z$-space; $z=0$ corresponds to the top edge of the top detectors. Since every source can be positioned separately they were simulated independently. To determine the optimum position for each source, only the relevant detector strings were taken into account: For source S1 string D1 was considered, for S2 the strings D2 and D3 and for source S3 the strings D2 and D4 (for the labels see figure~\ref{fig:p1top}). The single escape peak (SEP) of \power{208}Tl was chosen as reference peak because with an energy of $E = 2.104$\,MeV it is closest to the Q-value. Figure~\ref{fig:optpos} shows the number of counts in the SEP for each detector against the $z$ position. 

Two different cases were considered: 3 different $z$ positions per source, one for each detector of a string, and 2 different $z$ positions with the source between two detectors. To determine the best positions for both cases, a Gaussian was fitted to the count rates in each detector for the different $z$ positions (see figure~\ref{fig:optpos}). The peak positions of the Gaussians were used for the 3 position case, the intersection points for the 2 position case. Since the sources S2 and S3 have to calibrate two detector strings, the mean value between both strings was chosen. The only exception is the bottom detector in string D4: Since it is by far the smallest detector its optimum calibration position was weighted double when calculating the best position for source S3. The results are shown in table \ref{tab:optpos}.

To compare both cases, the total number of counts in the SEP for each detector were calculated using the optimum positions determined above and normalizing to a source activity of 20\,kBq per source and a calibration time of 1800\,s per position. The choice of these values will be explained in section~\ref{sec:activity}. The goal was to reach at least 1000\,counts in the SEP in each detector. Table~\ref{tab:countssep} shows the results for both cases. The different count rates per detector can be explained by their different masses (see figure~\ref{fig:optpos} where the normalization was done). Since two calibration positions are sufficient and more time efficient, these positions are used for the actual calibration\cite{gerdaOverview, gerda2nu2013}.

\subsection{Activity}\label{sec:activity}
In the next step the minimum source strength necessary to get sufficient statistics in all detectors has to be determined. Sufficient is defined in this case as a minimum of 1000\,counts in the peak as well as a peak to background ratio of 2\,:\,1. These values should be reached using the lowest possible activity within a total calibration run time of less than 4\,h including the movement of the sources. As a conservative estimate, a total moving time of 1\,h was assumed for lowering the source 5-6\,m from their parking position on top of the cryostat down to their two calibration positions and lifting them back afterwards. This leaves in total 3\,h for the calibration itself meaning 1.5\,h for each position. Since the half life of \tho\ is 1.9\,yr and the sources are left inside the experimental setup for the full Phase I ($1-2$\,yr including commissioning) a calibration time of 30\,min per position at the beginning of Phase I was assumed to ensure sufficient statistics also towards the end of this phase.

To estimate the necessary source activity the results of Monte Carlo simulations with each calibration source in its best positions were combined, therefore simulating a full calibration run. Again, the SEP was chosen as the reference peak and the number of counts as well as the peak to background ratio was determined for each detector. Since the third detector in string D4 (D4.3) has by far the smallest count rate, it was used as a reference.

\begin{figure}[t]
	\centering
	\subfigure[Total energy spectrum]{
		\includegraphics[width=.45\textwidth]{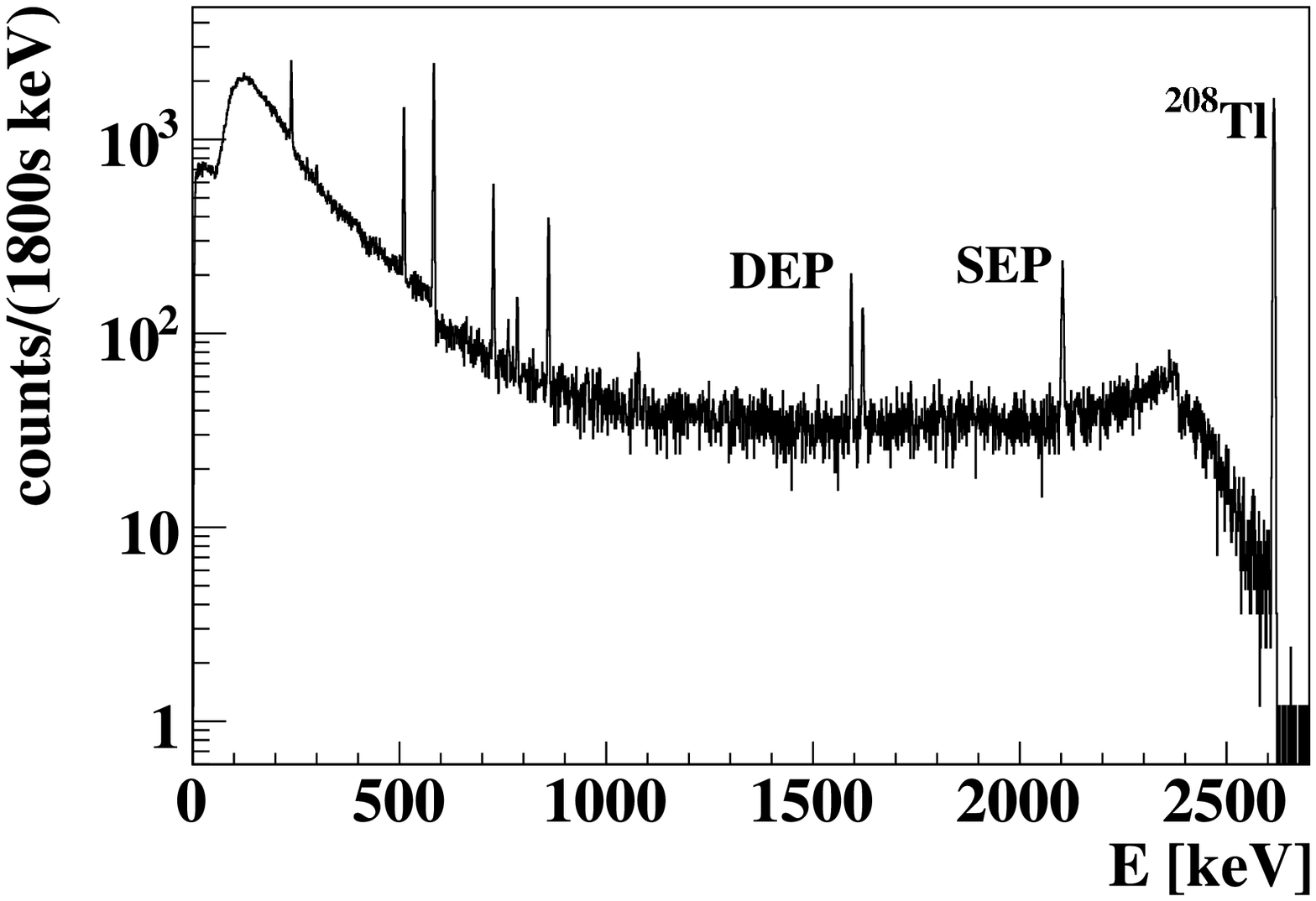}}
	\subfigure[Zoom into the region of interest] {
		\includegraphics[width=.45\textwidth]{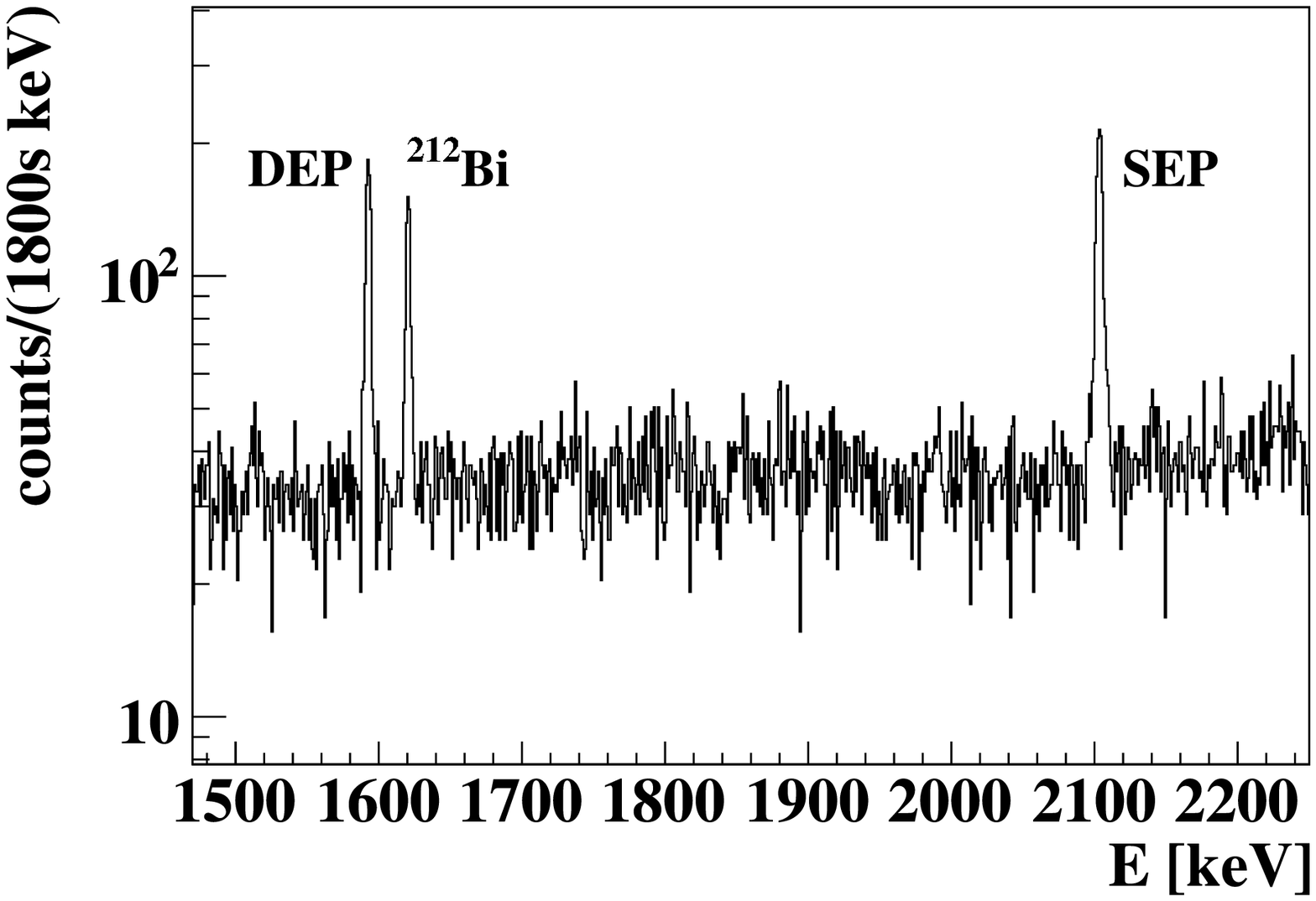}}
	\caption{Simulated energy spectrum of the smallest detector in the array after a calibration run with 3 sources with an activity of 20\,kBq each, 2 different $z$ positions and a run time of 30\,min per position.}
	\label{fig:ang1spec}
\end{figure}

A simulation of $3\times10^7$ decays per calibration source and position leads to 950\,counts in the SEP in detector D4.3 with a peak to background ratio of 2.4:1. Scaling these numbers up to 1000\,counts was considered as sufficient, leading together with the calibration time of 1800\,s to a minimum activity of $17.6\pm0.6$\,kBq. Therefore it was decided to use sources with an activity of $A=20$\,kBq for Phase I. Figure~\ref{fig:ang1spec} shows the energy spectrum of the smallest detector corresponding to this configuration.

\section{Background Contribution}
\label{sec:background}

Since a low background environment is crucial for the success of the experiment, an estimate of the background contribution in the ROI from the calibration system in the parking position is essential. Several isotopes in the \tho\ decay chain decay via $\alpha$ decay; hence it was considered both, the emitted $\gamma$'s and the neutrons produced via ($\alpha$,n) reactions in the surrounding material. The background contribution from neutrons during physics runs as well as calibration runs will be determined.

\subsection{Gamma Background}

\begin{figure}[t]
	\centering
	\subfigure[Full energy spectrum]{
		\includegraphics[width=.45\textwidth]{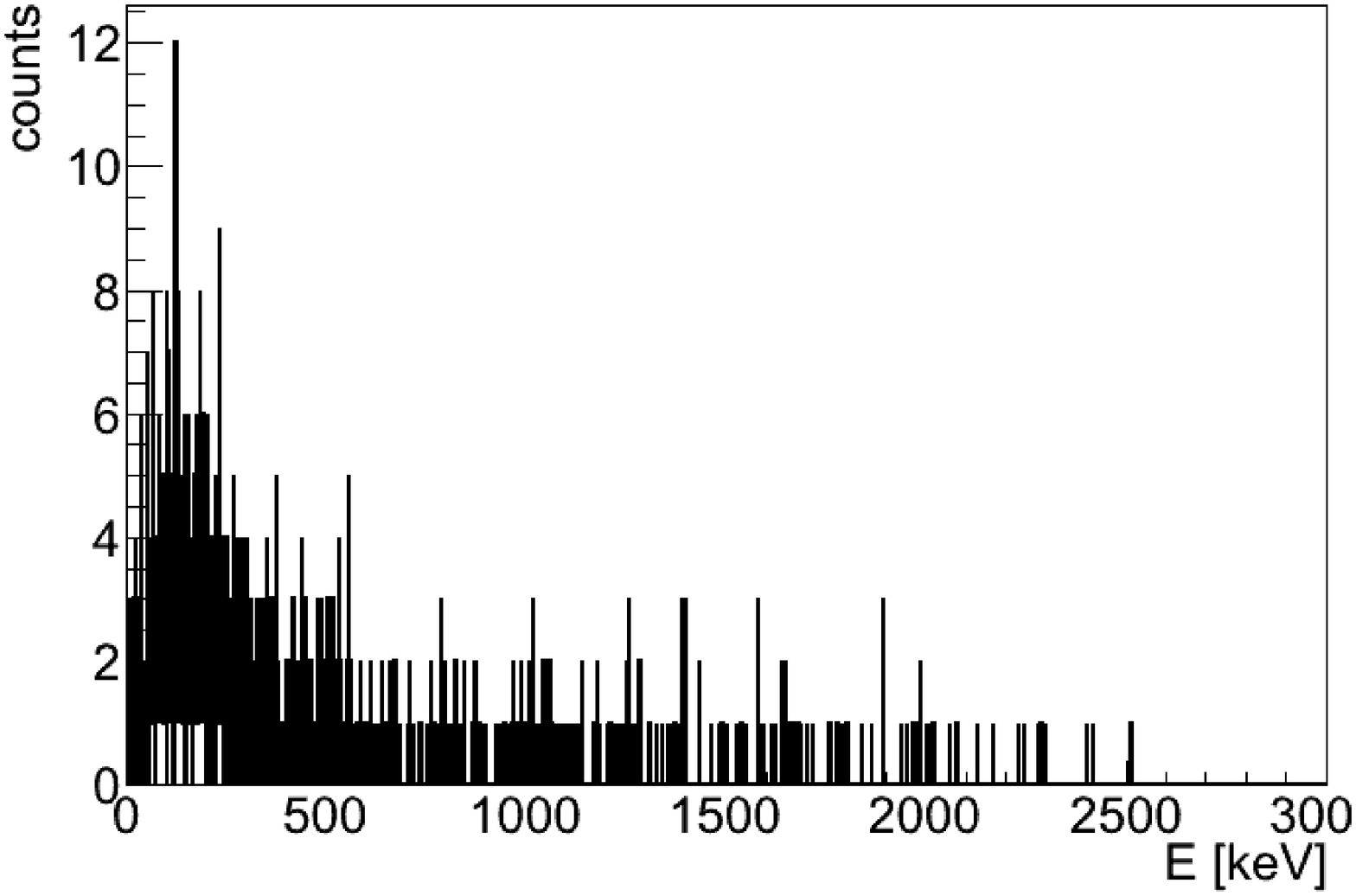}}
	\subfigure[Energy spectrum in the ROI] {
		\includegraphics[width=.45\textwidth]{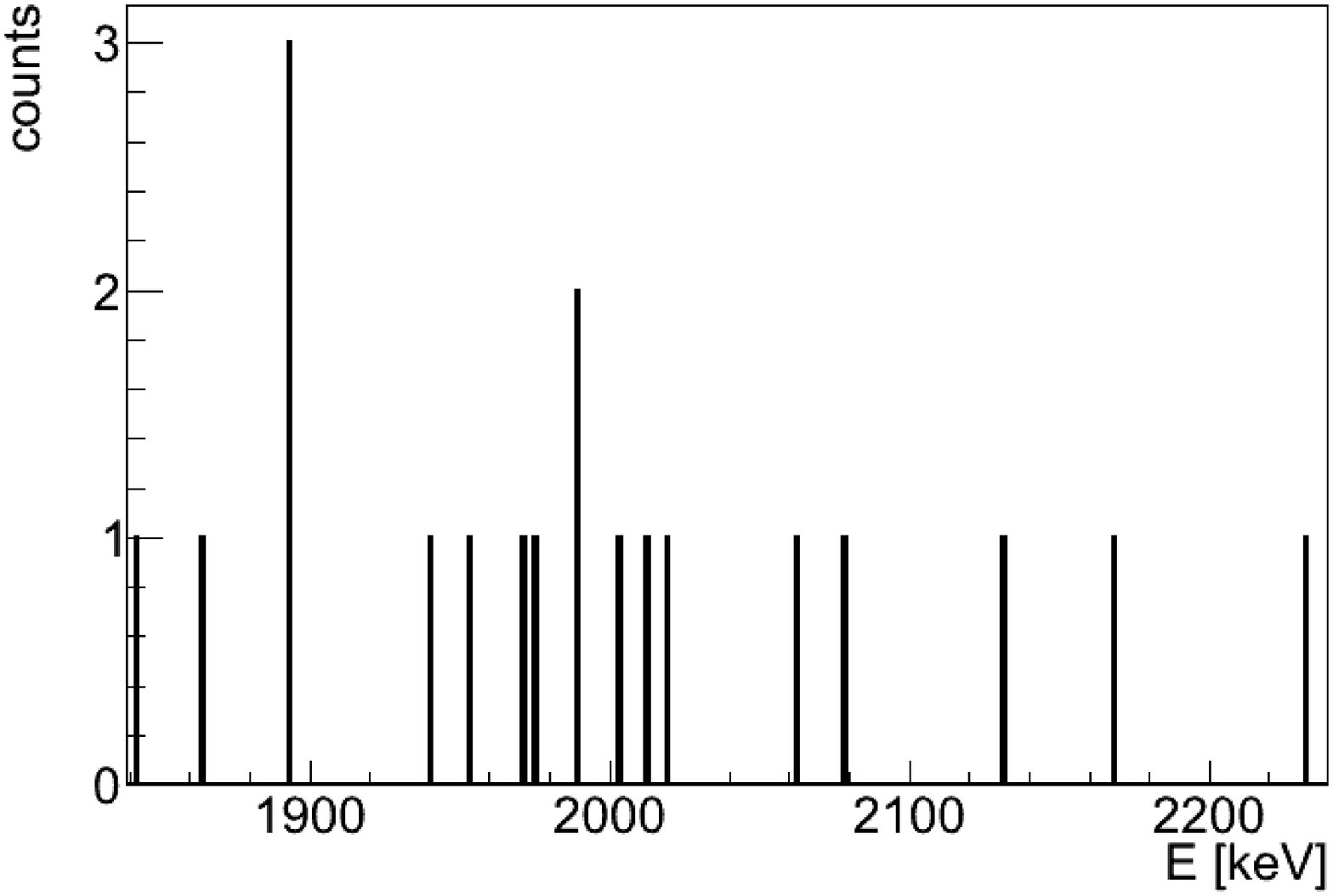}}
	\caption{Energy deposition of 2.6~MeV $\gamma$'s started 3.4~m above the detector array. A total of $1.05\times 10^{12}$ $\gamma$'s were simulated in a $2\pi$ hemisphere which corresponds to the number of \tho\ decays in about 3~years.}
	\label{fig:MC:mcback}
\end{figure}

According to original planing, there are a minimum of 3.4~m of liquid argon between the sources and the top detectors; the rest is gas and will not be considered here. In case of the $\gamma$ background, two methods were used and compared. The first one is a combination of an analytical estimate with Monte Carlo simulations (MCS), the second one is a pure MCS with the sources in their parking position. The latter is very CPU consuming. Therefore the first method was used for faster results to be able to determine the necessary shielding. The full MCS were used to verify the results.
\\\indent
In the semi-analytical method, MCS are used for the last 1~m of LAr. The flux at this position is estimated using linear attenuation $\phi = \phi_0\times e^{-d/l}$ with $\phi_0$ being the initial flux, $d$ the thickness of the absorbing material, $l=\frac{1}{\mu\rho}$ the mean free path, $\mu$ the mass attenuation coefficient and $\rho=1.394$\,g/cm$^3$ the density of liquid argon. The 2.6 MeV line from \power{208}Tl is the only line above the $Q$-value and therefore the only possible source for background in the ROI. In the following, only photons with this energy will be considered. The NIST database provides experimentally determined values for the mass attenuation coefficient for 2.044~MeV and 3.0~MeV photons; the value for the 2.6~MeV photons is interpolated according to their fit function. A mass attenuation coefficient of $\mu=0.0359$\,cm$^2$/g was found~\cite{berger10}, resulting in a mean free path of $l_\text{NIST}=20.0\pm0.4$\,cm. 
\\\indent
To confirm this result within the \gerda\ geometry, MCS were used: A beam of $10^8~\gamma$'s with $E=2.6$~MeV was directed to one detector with varying distances between them. Fitting an exponential decay to the total counts in the detector over distance lead to a mean free path of $l_\text{MCS}=20.65 \pm 0.05$\,cm. The corresponding plot can be found in figure~\ref{fig:disttest}. With this method, the mean free path could be determined with much higher precision than from the NIST database. This is because particle transportation and interaction where carried out step by step in a setting reflecting the actual situation in the \gerda\ cryostat. Since it is also more conservative, the mean free path determined by MCS will be used in the following.
\\\indent
For the initial flux, the number of decays from 3 calibration sources with $A=20$\,kBq in one year was used and then taken into account that due to the branching ratio just 36\% of these decays end with an emission of a 2.6\,MeV $\gamma$. Since the source radiates isotropically but the detector array with a radius of $r_{\text{det}} =15$\,cm covers only a small area of this sphere, this flux reduces by a factor of $\Omega_{\text{sphere}} / \Omega_{\text{det}} = 2055$, resulting in $\phi_0 = 3.3\times10^8\,\gamma$/yr. After 2.4\,m of LAr, the flux reduces to $\phi = 3.0 \times 10^3\,\gamma/$yr.
\\\indent
To obtain a conservative estimate of the background in the ROI, a photon beam directed to the center of one of the detector strings was simulated. Figure~\ref{fig:p1top} shows the $xy$ position of the detector strings and the calibration sources. S1 pointed to D1, S2 to D2 and S3 to D4. In total, $4.8\times 10^9$ $\gamma$'s with an energy of 2.6\,MeV were simulated. These simulations result in a total of $8.4\times 10^4$ events in the ROI. Rescaling to the initial flux and using a total mass of 17.7\,kg, a background index of \linebreak$(2.9 \pm 0.1 (\text{stat}) \pm 0.3(\text{sys}))\times 10^{-4}$\,\cts\ was obtained. Such a level is tolerable for Phase I with a background goal of $10^{-2}$~\cts. However, since other parts of the experiment, especially the cryostat, will contribute to the background as well, further shielding is preferred and will be discussed later.
\\\indent
In the full MCS three calibration sources were placed 3.4~m above the detector array. To acquire more statistics in a shorter time, more detector strings were included in the simulations leading to a total of 48 detectors with a total mass of 76.8~kg. Again, only 2.6~MeV $\gamma$'s are simulated since they are the only possible source for background in the ROI. Photons emitted in the top 2$\pi$ hemisphere will loose too much energy on their way to the detectors if they reach it at all and can be ignored. Therefore, just photons emitted in the lower 2$\pi$ hemisphere were simulated. The simulation of $1.05\times 10^{12}$ $\gamma$'s lead to a total 19~events in a 400 keV ROI from 1839-2239~keV. This large region was necessary due to the low statistics. The energy spectrum in the full energy range as well as the ROI is shown in figure~\ref{fig:MC:mcback}. This corresponds to a background contribution of $(2.0 \pm 0.6(\text{stat}) \pm 0.2(\text{sys}) )\times 10^{-4}$~\cts. As expected, the semi-analytical approach shows the more conservative limit but both values agree within errors.
\\\indent
To further shield the sources, the possible shielding material needs to be as radio-pure as possible with very good absorption properties and has to be machinable. The material fulfilling these requirements best is tantalum with $\mu (2.6\,\text{MeV}) = 0.04$\,cm$^2$/g, $\rho = 16.7$\,g/cm$^3$ and a natural radioactivity of about 50\,mBq/kg from \power{182}Ta with a half life of 114\,d. The detailed screening results measured with the Gator screening facility~\cite{gator} can be found in table~\ref{tab:screening}.
\begin{table}[t]
	\centering
	\caption{Screening results for tantalum}
	\begin{tabular}{r|cccccc}
	Isotope & \power{238}U & \power{232}Th & \power{60}Co & \power{40}K & \power{137}Cs & \power{182}Ta\\\hline
	Activity [mBq/kg]& $< 11$ & $< 9$ & $< 1.9$ & $< 33$ & $< 2.5$ & $52\pm5$
	\end{tabular}
	\label{tab:screening}
\end{table}
\\\indent
A reduction of the $\gamma$ background to $B_\gamma < 10^{-5}$\,\cts) was considered necessary. Using again linear attenuation to determine the needed minimum thickness of the shielding material, $d>5$\,cm was found. It was therefore decided to shield the calibration sources with 6\,cm of tantalum to be conservative, leading to a background contribution due to the $\gamma$ radiation of the calibration sources of
\begin{equation}
B_\gamma = (4.8 \pm0.1 (\text{stat})\pm0.3 (\text{sys}))\times 10^{-6}\text{ counts/(keV\cd kg\cd yr)}
\end{equation}

\subsection{Neutron background}
\label{sec:nback}
\begin{figure}[t]
	\centering
	\includegraphics[width=.4\textwidth]{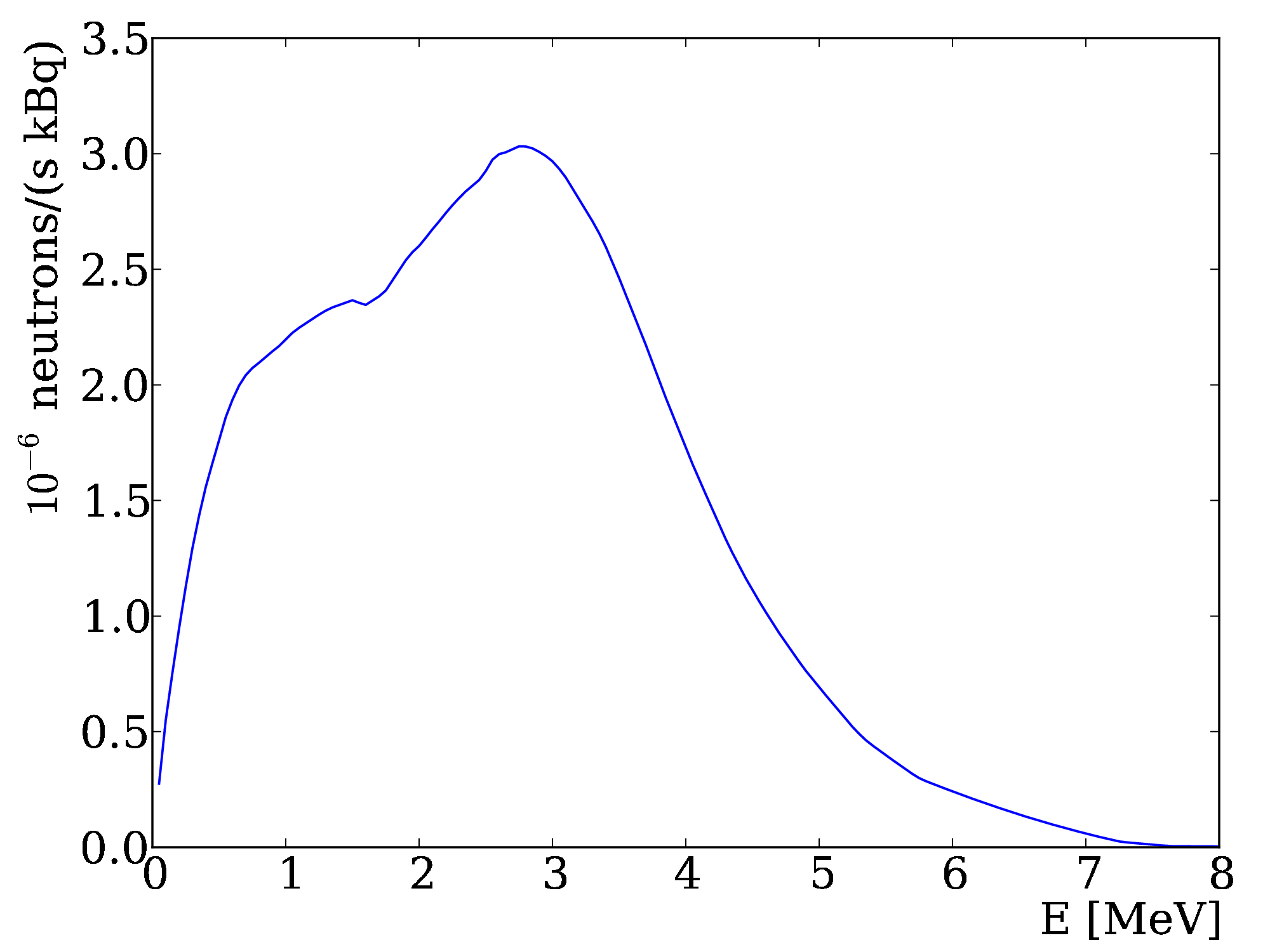}
	\caption{Predicted neutron spectrum from ($\alpha$,n) reactions of a \tho\ source embedded in gold.}
	\label{fig:nspec}
\end{figure}

Neutrons from ($\alpha$,n) reactions in the calibration source can contribute to the background in two different ways: By (n,n\textquotesingle$\gamma$) inelastic scattering of fast neutrons in the surrounding materials, especially in argon and due to neutron capture resulting in radioactive isotopes emitting photons with an energy close to the ROI. Elastic scattering is in this case not important since the maximum energy which can be transferred is far below the $Q$-value of 2039\,keV. For the scattering, the contribution from the parking position is most important. For the case of neutron capture, the calibration runs are also important to consider, since isotopes close or in the detectors might be activated. Therefore Monte Carlo simulations for both cases were performed. 

The neutron spectrum was determined using SOURCES4A~\cite{sources4a} with the following assumptions: A ThO$_2$ solution is coated on a gold foil placed in a stainless steel capsule~\cite{quellpaper}. Due to the high ($\alpha$,n) threshold of gold, the only reaction partners for the $\alpha$'s from \tho\ is oxygen which was taken into account in natural isotopic abundance in the calculations. The resulting neutron spectrum is shown in figure~\ref{fig:nspec}.

\begin{figure}[b]
	\centering
	\subfigure[Energy spectrum during a calibration run: The $\gamma$ contribution (black) is shown together with the background from elastric scattering of neutrons in the detectors as well as (n,n\textquotesingle)$\gamma$'s (red).] {
	\includegraphics[width=.31\textwidth]{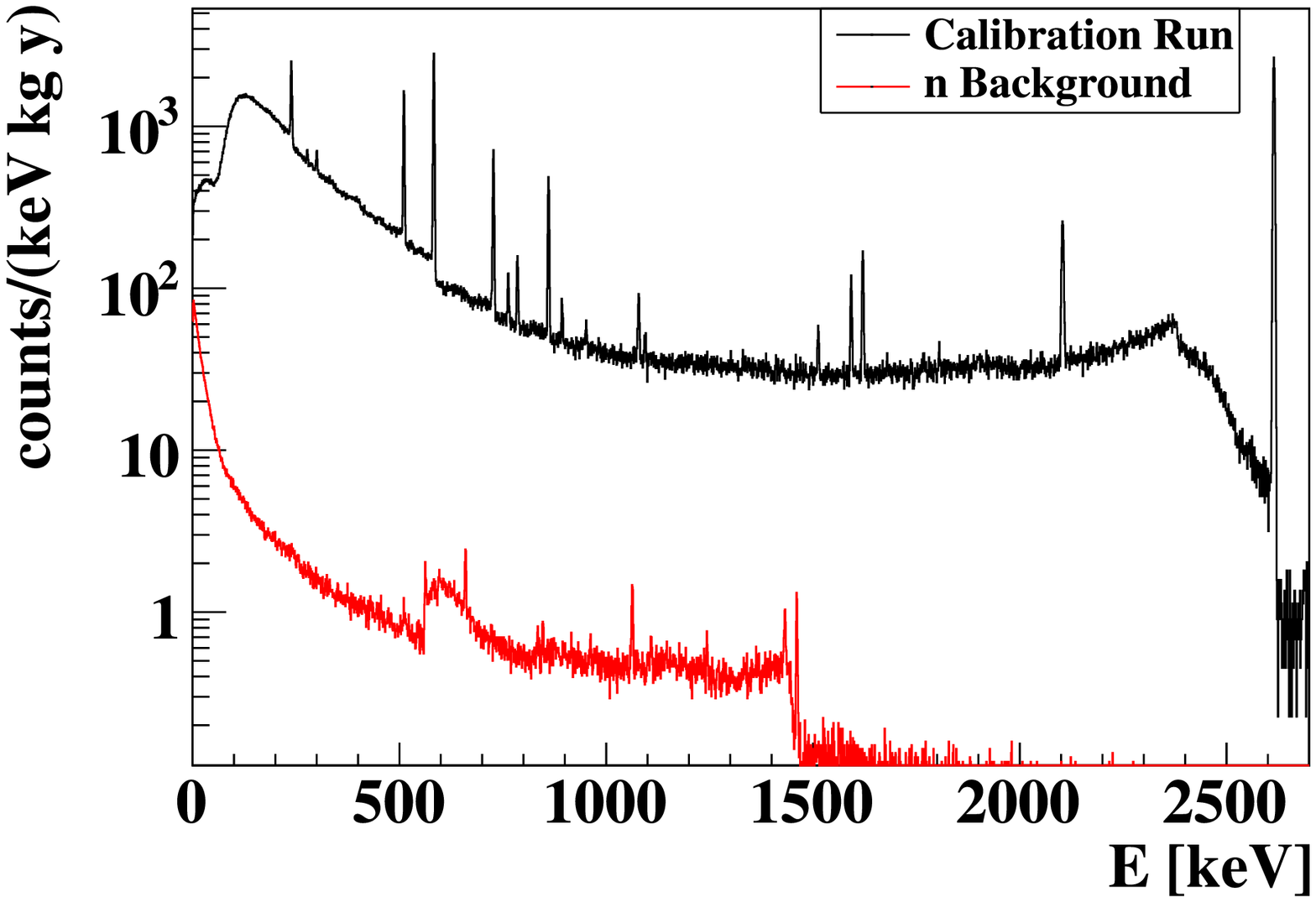}
	\label{fig:calrun_nback}}\hspace*{1mm}
	\subfigure[Background contribution from the sources in parking position due to neutron scattering events.] {
	\includegraphics[width=.31\textwidth]{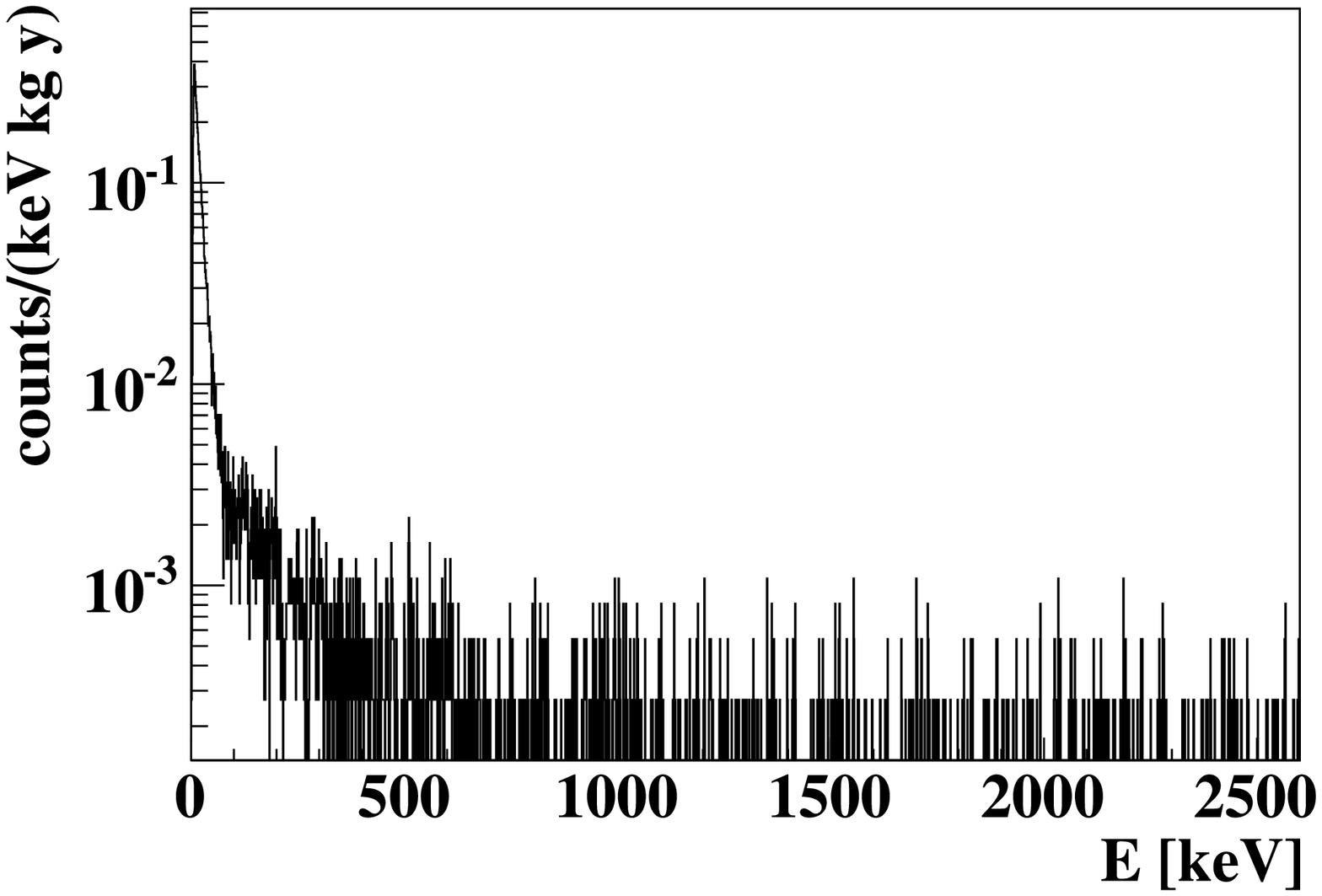}
	\label{fig:nback_scatter}}\hspace*{1mm}
	\subfigure[Background contribution from the sources in parking position due to \power{77}Ge produced via neutron capture.] {
	\includegraphics[width=.31\textwidth]{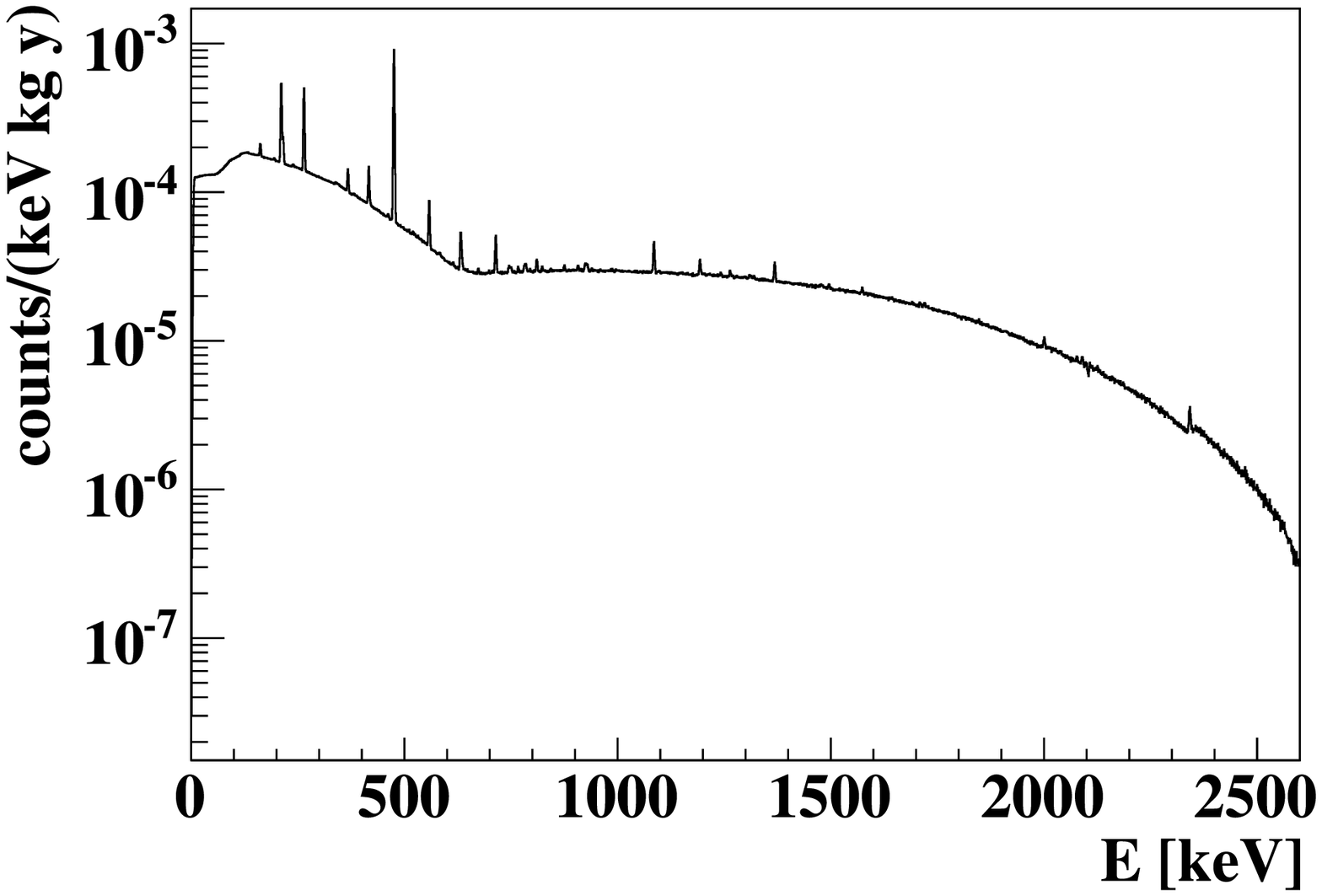}
	\label{fig:nback_ge77}}
	\caption{Neutron background during calibration (a) and from parking position (b) and (c).}
	\label{fig:calvsnback}
\end{figure}

As a first step, the background contribution from elastic scattering of neutrons in the detectors as well as (n,n\textquotesingle)$\gamma$'s during the calibration run was inspected. As figure~\ref{fig:calvsnback} shows, this background has no significant influence on the calibration spectrum. Therefore it will be ignored in the following.

For the estimation of the background due to neutron scattering, Monte Carlo simulations were started with the sources in their parking position and the neutron spectrum shown in figure~\ref{fig:nspec}. The simulation of $10^9$ neutrons resulted in 230\,counts in the energy region between [1950,2060]\,keV. As a realistic estimate, a neutron flux of $A_\text{n} = 10^{-3}$~n/(s\cd kBq)~\cite{quellpaper} was assumed. Using these numbers for three calibration sources with an activity of $A=20$\,kBq each the following background contribution was found:
\begin{equation}
B_\text{n}^\text{scatter} = (6.2\pm0.4(\text{stat})) \times 10^{-5} \pm 20\%(\text{sys)\,\cts}
\end{equation}
which is sufficiently low for \gerda\ Phase I.

\begin{figure}[t]
	\hspace*{-2mm}
	\subfigure[Activated isotopes during calibration.]{
		\includegraphics[width=.49\textwidth]{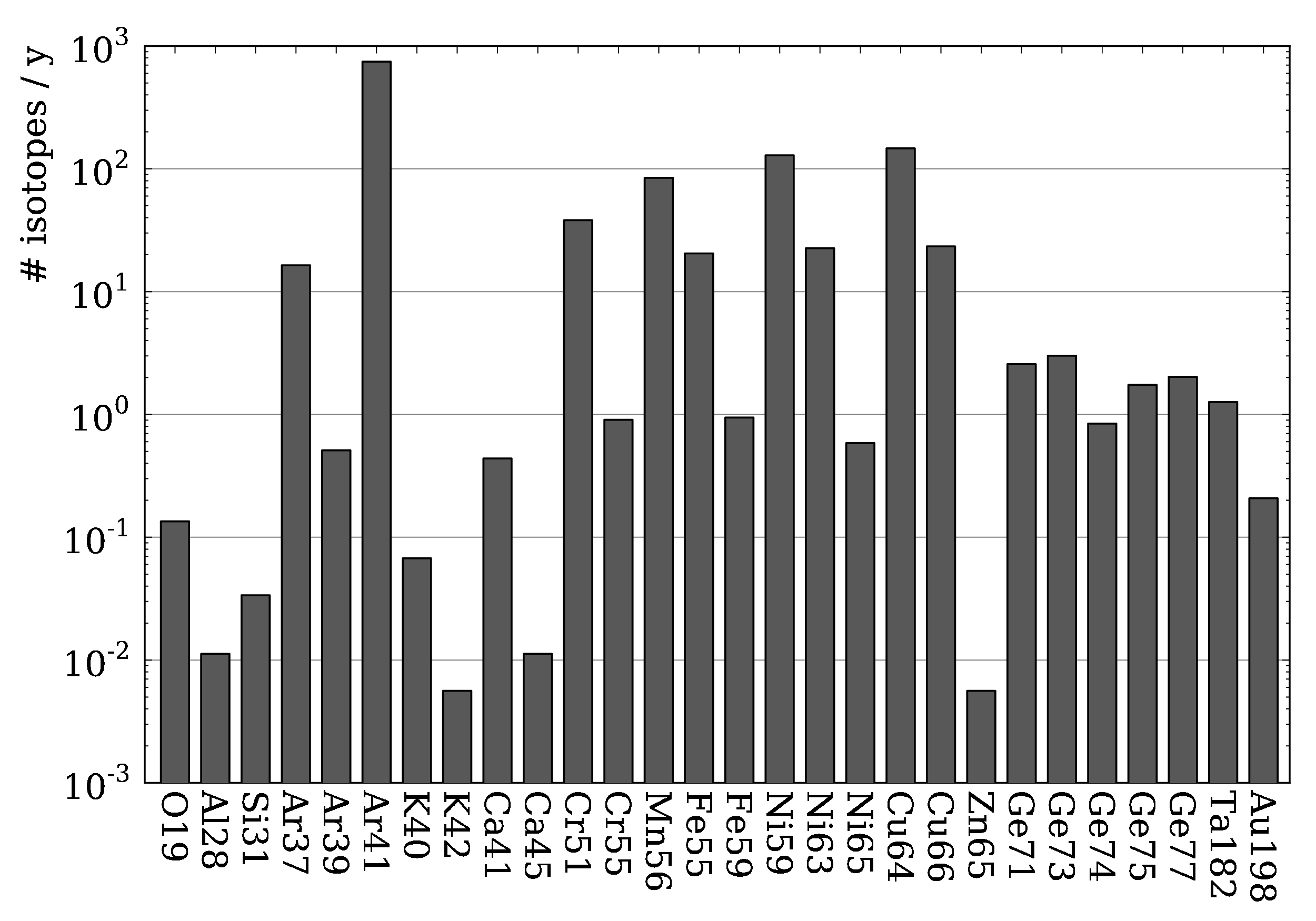}
		\label{fig:actIsoCal}
	}\hspace*{0mm}
	\subfigure[Activated isotopes from source in parking position.]{
		\includegraphics[width=.49\textwidth]{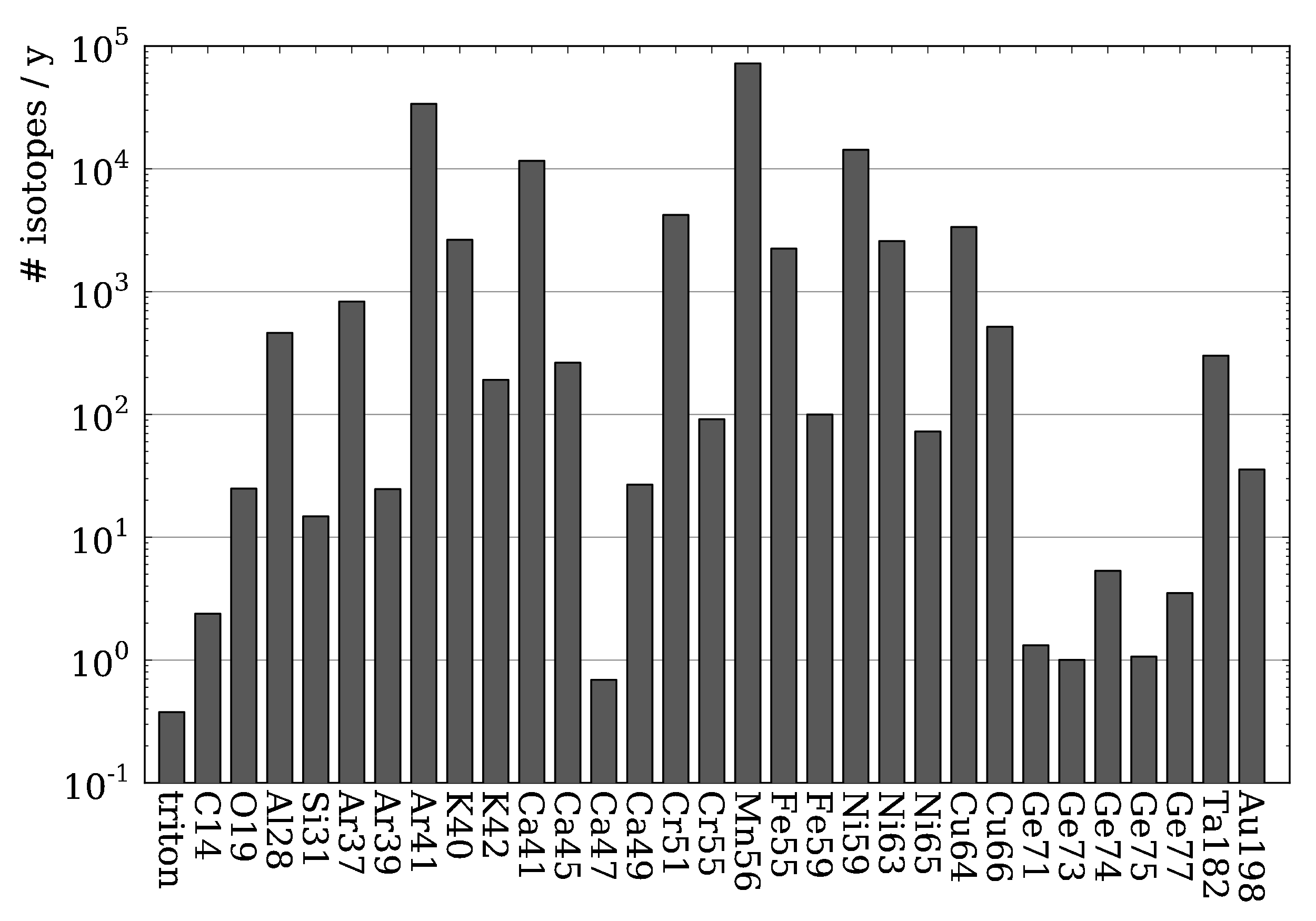}
		\label{fig:actIsoPark}
	}
	\caption{Activated isotopes normalized to amount produced per year.}
	\label{fig:actIso}
\end{figure}

In the next step the isotopes produced due to neutron capture reactions in all parts of the experiments were analyzed. Figure~\ref{fig:actIso} shows the type and amount of produced isotopes in one year during calibration runs (figure~\ref{fig:actIsoCal}) as well as from the parking position (figure~\ref{fig:actIsoPark}), assuming one calibration run of 1\,h per week. Most of these isotopes do not emit photons nor $\beta$'s with energies above 2\,MeV and are thus no potential background candidates. Produced isotopes emitting photons or $\beta$'s above 2\,MeV are: \power{19}O, \power{28}Al, \power{41}Ar, \power{42}K, \power{49}Ca, \power{55}Cr, \power{56}Mn, \power{66}Cu and \power{77}Ge. 

Concerning the photon emitters, those isotope, whose intensity is below 1\,\%, were ignored. This reduces the possible background candidates to \power{49}Ca, \power{56}Mn and \power{77}Ge. \power{56}Mn is produced in several volumes, the highest production rates are the rock of the laboratory ($6.2\times10^4$\,isotopes/yr) and the wall of the cryostat ($2\times10^4$\,isotopes/yr), assuming again a neutron flux of $A_n = 10^{-3}$\,n/(s\cd kBq) and a total activity of \linebreak$3\times20$\,kBq. Due to the continuous neutron flux from the sources in parking position and the short half-life of \power{56}Mn of 2.6\,h, the production and decay rate are in equilibrium, leading to an activity in the rock of $2.8\times10^{-4}$\,Bq and in the wall of the cryostat of $8.9\times10^{-5}$\,Bq. A total of $1.2\times10^9$\,decays were simulated in the cryostat wall. No events were found above 1.5 MeV leading to an upper limit on the background contribution of \linebreak$<1.5\times10^{-9}$\,\cts, which is negligible. Since the water tank further shields the detectors from the radiation of the rock, its contribution can be ignored. The same argument holds for \power{49}Ca produced also in the rock of the laboratory with a very low production rate of 26 isotopes/yr.

Due to the short stopping range of $\beta$'s, only isotopes produced inside the cryostat were considered. This reduces the possible background candidates to \power{41}Ar and \power{77}Ge. \power{41}Ar, with a half life of 109\,m and a Q-value for the $\beta$-decay of 2.5\,MeV, is produced in the liquid argon with a production rate of $2.6\times10^6$\,isotopes/yr, assuming again a neutron flux of $A_n = 10^{-3}$ n/(s\cd kBq) and a total activity of $3\times20$\,kBq. A total of $2.4\times10^9$\,decays were simulated, isotropically distributed in the cryo liquid. To achieve reasonable statistics, the 200\,keV interval around the Q-value was chosen as ROI. A total of 48\,counts were found, leading to a negligible background contribution of $(7.3\pm 1.1)\times10^{-8}$\,\cts.

Therefore, only \power{77}Ge, produced in the detectors themselves, is potentially dangerous. After the neutron capture, the \power{77}Ge is highly excited and de-excites via a $\gamma$ cascade into either the ground state or the metastable state \power{77m}Ge. The ground state $\beta$ decays with $T_{1/2} = 11.3$\,h, with endpoint energies up to 2.5\,MeV and emits several gammas with energies up to 2.3\,MeV. \power{77m}Ge decays with $T_{1/2} = 53$\,s with a 19\,\% chance into the ground state emitting a 160\,keV $\gamma$ or with a 81\,\% chance via $\beta$ decay with endpoint energies up to 2.9\,MeV, emitting several gammas with energies up to 1.7\,MeV. The total production rate of both, \power{77}Ge and \power{77m}Ge, is $A_{^{77}\text{Ge}} = 8.6$\,isotopes/yr, assumed to be equivalent to the decay rate. Since the corresponding cross sections vary significantly~\cite{lyon57, mateosian57, meierhofer09}, the range of the background contribution was estimated using two simulations: One assuming that the total amount of the produced \power{77}Ge will decay from the ground state and the other one from the metastable state. In both cases a total of $2\times10^7$ decays were simulated, resulting in $B_\text{ground} = 2.9\times10^{-5}$\,\cts and $B_\text{meta} = 4.0\times10^{-5}$\,\cts. As a conservative limit, $B_\text{ground}$ will be used:
\begin{equation}
	B_{^{77}\text{Ge}} = (4.0\pm0.1(\text{stat})\,^{+0.3}_{-1.4}(\text{sys})\,)\times10^{-5}\text{ counts/(keV\cd kg\cd y)}
\end{equation}
Combining the different background contributions due to neutrons,
\begin{equation}
	B_\text{n} = (1.0\pm 0.1(\text{stat})\,^{+0.1}_{-0.2}(\text{sys}))\times 10^{-4}\,\text{counts/(keV\cd kg\cd y)}
\end{equation}
was found, which is well below the \gerda\ Phase I goal but relatively close to the Phase II goal. Thus, further reduction is preferred for this later stage of the experiment.


\section{Conclusion}
\label{sec:conclusion}
The Monte Carlo studies for the calibration system of the \gerda\ experiment showed that three \tho\ sources with an activity of 20\,kBq each are necessary to calibrate the detector array in Phase I of the experiment. To prevent scattering of the $\gamma$'s in the liquid argon which is used to cool and shield the detectors, the calibration sources were placed in the horizontal plane as close as possible to the detectors. In the vertical direction two positions between the detector layers are necessary to reach sufficient statistics in each detector within a calibration time of 30~min.

During a physics run the calibration sources are parked on top of the cryostat and the radiation of the sources might contribute to the background in the region of interest. Both gamma as well as neutron radiation as result of ($\alpha$,n) reactions were considered and a total background contribution of $(1.07\pm0.04(\text{stat})^{+0.13}_{-0.19}(\text{sys}))\times 10^{-4}$\,\cts\ was found. This is well below the background goal for Phase I of $10^{-2}$\,\cts\ but might become more relevant for Phase II with a background goal of  $10^{-3}$\,\cts. The highest contribution is due to ($\alpha$,n) neutrons, which make 93\% of the total expected background. A shielding of each source with 6\,cm of tantalum was necessary to reach such a low background contribution.


\section{Acknowledgements}
This work was supported by the University of Zurich, the Swiss National Foundation grant 200020\_138225 and the Friedrich Ebert Foundation.

\bibliographystyle{model1-num-names.bst}
\bibliography{./Literatur.bib}

\end{document}